\documentclass[%
reprint,
 amsmath,amssymb,
 aps,
floatfix,
]{revtex4-2}

\usepackage{soul}
\usepackage{float}
\usepackage{graphicx}
\usepackage{dcolumn}
\usepackage{bm}
\usepackage{hyperref}
\usepackage[caption=false]{subfig}

\usepackage{hyperref}
\hypersetup{
    colorlinks=true,
    linkcolor=blue,
    filecolor=magenta,      
    urlcolor=cyan,
    anchorcolor=blue,
    citecolor=blue,
    linkcolor=blue,
    menucolor=blue,
}

\usepackage{subfiles}
\def\biblio{\bibliography{reference}}

\usepackage{tikz}
\usetikzlibrary{calc,decorations.markings,intersections}
\usepackage{upgreek}
\usepackage{pgfplots}
\pgfplotsset{width=\linewidth,compat=1.9}

\begin{document}
\def\biblio{}

\preprint{APS/123-QED}

\title{Acoustic detection potential of single minimum ionizing particles in viscous liquids}

\author{Panagiotis Oikonomou}
    \email{panoso@nyu.edu}
\author{Laura Manenti}
    \email{laura.manenti@nyu.edu}
\author{Isaac Sarnoff}
\author{Francesco Arneodo}
    \email{francesco.arneodo@nyu.edu}
\affiliation{Division of Science, New York University Abu Dhabi, United Arab Emirates}
\affiliation{Center for Astro, Particle and Planetary Physics (CAP$^3$), New York University Abu Dhabi, United Arab Emirates}

\date{\today}

\begin{abstract}
An ionizing particle passing through a liquid generates acoustic signals via local heat deposition. We delve into modeling such acoustic signals in the case of a single particle that interacts with the liquid electromagnetically in a generic way. We present a systematic way of introducing corrections due to viscosity using a perturbative approach so that our solution is valid at large distances from the interaction point. A computational simulation framework to perform the calculations described is also provided. The methodology developed is then applied to predict the acoustic signal of relativistic muons in various liquids as a toy model.
\end{abstract}

\maketitle

\tableofcontents


\section{Introduction\label{sec:introduction}}

Research in the production of acoustic signals by charged particles in liquids received increasing attention after the characterization of acoustic waves produced by ``local heating'' in 1957~\cite{Askaryan1957}. Charged particles passing through liquids give rise to acoustic radiation primarily via local heating~\cite{Lahman2015}, if the energy deposited is large enough, the acoustic wave becomes detectable. Hence, acoustic signals produced by heavy ion beams, PeV particle cascades, and high energy neutrinos were heavily investigated between the 1960s and the early 1990s~\cite{Roberts1992,Karle2003,Askaryan1979,Sulak1979,Jiang1985,Bowen1979,Albul2001}.

Acoustic detection has traditionally been used as a tool to detect high energy cosmic particles~\cite{Budnev2017,Ishihara2019,Nosengo2009}. However, it has recently found applications at lower energies in the search for cold dark matter using superheated liquids~\cite{Baxter2017,Barnabe2005,Behnke2008,Amole2019}, where ionizing particles produce a shockwave during a local, instantaneous phase transition of the liquid~\cite{Amole2015}.

Such detection technique comes at the cost of conducting the experiment in a specialized thermodynamic state so that nonlinear effects, such as molecular dissociation, microbubble formation, shocks, etc., are present~\cite{Hunter1981}.

In a liquid that is not in a finely controlled thermodynamic state, the theoretical calculation of the acoustic signal produced by low energy particles ($\mathcal{O}(\rm keV)$) is non-trivial. Even in ideally static fluids, current theoretical solutions break down at large distances from the interaction point (e.g. because the viscosity of the liquid is not taken into account~\cite{Learned1979}).

In this manuscript, we show how to estimate acoustic signals produced in liquids by a single ionizing particle interaction in the most generic and accurate method to date.

Instead of specializing in local heating, we consider the more general case of a moving, compactly supported heat deposition of arbitrary shape. We also introduce a way to account for the viscosity of the liquid so that second order corrections can be easily calculated, allowing for a more realistic solution of the acoustic wave at large distances from the particle track, where the signal is highly attenuated. 
To make things painless for our reader, we explicitly derive all equations, always clearly stating the physical assumptions behind each mathematical step. Finally, as an example, we present the calculation of the acoustic signal of relativistic muons using our newly developed methodology.

The study presented can also be used as a resource for the willful experimental physicist that embarks on the mission of detecting acoustic signals of particles in viscous liquids. 

\biblio



\section{Strongly Damped Acoustic Wave Equation\label{sec:derivations}}

We introduce the mechanism behind the generation of an acoustic signal due to a single particle interaction in a liquid. Let us first consider the effect of an arbitrary heat deposition in the bulk of the liquid. For this purpose, we add a source term and a damping term to the well-known acoustic wave equation in an isothermal fluid:

\begin{equation}
    \Delta p(\bm{x},t) = \rho_0 \kappa \frac{\partial^2 }{\partial t^2} p(\bm{x},t),
    \label{eq:wave-eq}
\end{equation}

where $p(\bm{x},t)$ is the pressure difference at a location $\bm{x}$ in the liquid at time $t$, $\rho_0$ is the rest density taken as constant, and $\kappa$ is the compressibility of the liquid.

\subsection{Damping Term Derivation \label{subsec:damping-term-derivation}}

\subsubsection{Derivation from Navier Stokes}

Due to the small energy deposition by a single particle in the fluid, it is reasonable to assume that the damping effect caused by the viscosity of the liquid will significantly affect the decay time of the acoustic wave. To derive the viscous wave equation (also known as the strongly damped wave equation \cite{Bortolan2015,Viggen2011}), we exploit the principle of conservation of mass and momentum expressed by Eq.~\eqref{eq:mass-conservation} and Eq.~\eqref{eq:momentum-conservation}, and include the damping term $\mu \Delta \bm{v}$ in the latter:
\begin{align}
    \frac{\partial \rho}{\partial t} + \bm{\nabla} \cdot (\rho \bm{v}) &= 0 \label{eq:mass-conservation}\\
    \rho \frac{\partial \bm{v}}{\partial t} + \rho (\bm{v}\cdot \bm{\nabla}) \bm{v} &= -\bm{\nabla} P + \mu \Delta \bm{v}, \label{eq:momentum-conservation}
\end{align}
where $\rho(\bm{x},t)$, $P(\bm{x},t)$, and $\bm{v}(\bm{x},t)$ are the density, pressure, and velocity, respectively, while $\mu$ is the coefficient of bulk viscosity. In the same fashion as other related literature~\cite{Learned1979,Viggen2011,Settnes2012}, we assume a fluid with no vorticity ($\bm{\nabla} \times \bm{v} = \bm{0}$). Thus, Navier-Stokes equation Eq.~\eqref{eq:momentum-conservation} takes the form
\begin{equation}
    \rho \frac{\partial \bm{v}}{\partial t} + \rho \bm{v}\, \left(\bm{\nabla} \cdot \bm{v}\right) = -\bm{\nabla} P + \mu \Delta \bm{v}.
    \label{eq:navier-stokes}
\end{equation}
Let us now consider a small perturbation to each of the variables:
\begin{align}
    \rho &= \rho_0 + \delta \rho\\
    P &= p_0 + \delta p\\
    \bm{v} &= \bm{v}_0 + \delta \bm{v} = \delta \bm{v},
\end{align}
where  $\rho_0$, $p_0$, and $\bm{v}_0$ are the density, pressure, and velocity of the fluid at equilibrium, and we have assumed the fluid to be initially at rest. Since we will exclusively deal with $\delta p, \delta \rho$ and $\delta \bm{v}$, from now on we shall drop the delta to keep notation light. By plugging the perturbed variables into Eq.~\eqref{eq:mass-conservation} and Eq.~\eqref{eq:navier-stokes} and neglecting higher order terms, i.e. $\mathcal{O}((\delta \rho/\rho_0)^2)$, $\mathcal{O}((\delta P/P_0)^2)$, and $\mathcal{O}((\delta v/v_0)^2)$, we obtain the following equations:
\begin{align}
    \frac{\partial \rho}{\partial t} + \rho_0 \bm{\nabla} \cdot \bm{v} &= 0 \label{eq:mass-conservation-linear}\\
    \rho_0 \frac{\partial \bm{v}}{\partial t} + \bm{\nabla} p &= \mu \Delta \bm{v}. 
    \label{eq:navier-stokes-linear}
\end{align}
Taking the divergence of Eq.~\eqref{eq:navier-stokes-linear} leads to
\begin{equation}
    \rho_0 \frac{\partial}{\partial t} \bm{\nabla} \cdot \bm{v} + \Delta p = \mu \Delta \left(\bm{\nabla} \cdot \bm{v}\right),
    \label{eq:navier-stokes-linear-II}
\end{equation}
where $\bm{\nabla} \cdot \bm{v}$ can be replaced using Eq.~\eqref{eq:navier-stokes-linear-II} to obtain 
\begin{equation}
    \frac{\partial^2 \rho}{\partial t^2} = \Delta \left(p + \frac{\mu}{\rho_0} \frac{\partial \rho}{\partial t}\right).
    \label{eq:navier-stokes-linear-III}
\end{equation}
By writing the density as a function of the compressibility, i.e. $\rho = \rho_0 \kappa p$, Eq.~\eqref{eq:navier-stokes-linear-III} can be expressed only as a function of pressure:
\begin{equation}
    \Delta \left(p + \mu\, \kappa \frac{\partial p}{\partial t}\right) = \rho_0\, \kappa\, \frac{\partial^2 p}{\partial t^2}.
    \label{eq:wave-specific}
\end{equation}
We define the speed of the wave, $c$, as $c = 1/\sqrt{\rho_0 \kappa}$ and the attenuation frequency, $\omega_0$, as $\omega_0 = 1/\mu \kappa$. Thus, Eq.~\eqref{eq:wave-specific} takes the more familiar form:
\begin{equation}
    \Delta \left(p + \frac{1}{\omega_0} \frac{\partial p}{\partial t}\right) = \frac{1}{c^2} \frac{\partial^2 p}{\partial t^2}.
    \label{eq:wave}
\end{equation}

\subsubsection{Damping Characteristic}
We obtain the dispersion relation of plane wave solutions to Eq.~\eqref{eq:wave} by taking the Fourier transform of the pressure in space and time as defined later in Eq.~\eqref{eq:fourier}. By doing so, we obtain that the dispersion relation between the frequency $\omega$ and wavenumber $k = ||\vec{k}||$ is given by
\begin{equation}
    \label{eq:dispersion}
    k^2 = \frac{\omega^2}{c^2} \left(1-i\frac{\omega}{\omega_0}\right)^{-1}.
\end{equation}
In the limit of $\omega_0 \to \infty$ (i.e. in the case of no damping) Eq.~\eqref{eq:dispersion} becomes $kc = \omega$ which is the expected solution. 

Equation \eqref{eq:dispersion} also shows how the damping term affects the sound wave generated. At low frequency, the relation is similar to the undamped case ($\omega_0 \to \infty$). However, as the frequency increases, the wavenumber imaginary component further increases. 

We can understand the physical significance of the imaginary component of $k$ by plugging it into a plane wave $e^{i k(\omega) r - i \omega t}$. The imaginary part is going to introduce an exponential decay in the amplitude of the wave with a rate proportional to it and that increases with frequency. In the context of acoustic waves produced by particles, the higher frequency components of the generated pressure wave will decay faster, while the lower frequency components will propagate further, making it easier to detect them. 

\subsection{Source Term Derivation \label{subsec:source-term-derivation}}

It is well documented~\cite{Groom2021,Kutcher1976} that a particle passing through a liquid deposits energy such that the local temperature sharply increases. The almost instantaneous change in temperature leads to a rapid volume expansion, and the subsequent change in density propagates through the liquid. Here, we assume that this effect, referred to as ``local heating'', is the biggest contributor to the generation of the sound wave. This is consistent with past literature where acoustic signals due to particle beams were studied~\cite{Vandenbroucke2005,Learned1979,Askaryan1979}. In this section, we address the mechanism by which the wave is generated, i.e., the source term in Eq.~\eqref{eq:wave-eq}.

\subsubsection{Derivation from Navier Stokes}

Let us consider the effect of some local temperature fluctuation $\tau (\bm{x},t)$ such that the total temperature is given by $T(\bm{x},t) = T_0 + \tau(\bm{x},t)$, where $T_0$ is the equilibrium temperature.
With a variation in temperature, density will change as a function of both pressure and temperature. Specifically, at the first order (using $\rho,\ p$, and $\tau$ to denote the changes in density, pressure, and temperature, respectively) we can express the change in density by \cite{Morse1986}:
\begin{equation}
    \rho = \left.\frac{\partial \rho}{\partial P}\right|_T p + \left.\frac{\partial \rho}{\partial T}\right|_P \tau,
    \label{eq:density-change}
\end{equation}
where $P$ and $T$ represent the total pressure and temperature respectively, and $V$ represents a small volume of the liquid around the particle interaction point. Let us recall the definitions of isothermal compressibility $\kappa_T$ and coefficient of thermal expansion \cite{Lyamshev2004}:
\begin{align}
    \kappa_T &= -\frac{1}{V} \left. \frac{\partial V}{\partial P}\right|_T = \frac{1}{\rho_0} \left. \frac{\partial \rho}{\partial P}\right|_T
    \label{eq:compressibility}\\
    \beta &= \frac{1}{V} \left. \frac{\partial V}{\partial T}\right|_P = - \frac{1}{\rho_0} \left. \frac{\partial \rho}{\partial T}\right|_P.
    \label{eq:thermal-expansion}
\end{align}
We can then rewrite Eq.~\eqref{eq:density-change} as follows \cite{Morse1986}:
\begin{equation*}
    \rho = \rho_0 (\kappa_T p - \beta \tau).
\end{equation*}
From here we can proceed just as in Section~\ref{subsec:damping-term-derivation} when we derived Eq.~\eqref{eq:navier-stokes-linear-III}. First, we use Eq.~\eqref{eq:density-change} to write the density in terms of the pressure difference $p$ and temperature fluctuation $\tau$. Then, we assume that, to first order, the functions $\kappa_T(P,T,t)$ and $\beta(P,T,t)$ vary slowly with time, such that
\begin{equation}
    \rho_0 \kappa_T \frac{\partial^2 p}{\partial t^2} - \rho_0\beta\frac{\partial^2 \tau}{\partial t^2} = \Delta \left(p + \mu \kappa_T \frac{\partial p}{\partial t} + \mu \beta \frac{\partial \tau}{\partial t}\right).
    \label{eq:wave-source-I}
\end{equation}
Equation~\eqref{eq:wave-source-I} shows the presence of an extra damping term. However, assuming that the temperature variation is small,  $\frac{\partial}{\partial t} \Delta \tau \approx 0$, we may neglect it. Using the first law of thermodynamics, we can determine the heat per unit volume added to the liquid $\epsilon(\bm{x},t)$
\begin{equation}
    \epsilon = \frac{\delta Q}{\delta V} = \rho_0 C_p \tau,
    \label{eq:heat-per-volume}
\end{equation}
where $C_p$ is the specific heat capacity of the liquid at constant pressure. We may now introduce the complete wave equation by substituting $\tau$ with Eq.~\eqref{eq:heat-per-volume}:
\begin{equation}
      \Delta \left(p + \mu \kappa_T \frac{\partial p}{\partial t} \right) - \rho_0 \kappa_T \frac{\partial^2 p}{\partial t^2} = - \frac{\beta}{C_p} \frac{\partial^2 \epsilon}{\partial t^2}
    \label{eq:wave-source-spec}
\end{equation}
which may be more compactly written as
\begin{equation}
      \Delta \left(p + \frac{1}{\omega_0} p_t \right) - \frac{1}{c^2} p_{tt} = - \frac{\beta}{C_p} \epsilon_{tt}.
    \label{eq:wave-source}
\end{equation}
We have successfully derived the correction terms to the acoustic wave equation Eq.~\eqref{eq:wave-eq} that is generated by a heat source inside a liquid. In Section~\ref{subsec:energy-deposition}, we will show how to estimate the heat deposition $\epsilon(\bm{x},t)$ for a single, charged particle through a liquid using the Bethe-Bloch formula. At this stage it is important to note that our assumption that $\kappa$ and $\beta$ do not vary in time is only valid away from the point where the energy deposition occurs.

\biblio



\section{Analytic Solutions to the Wave Equation}
\label{sec:solutions}

In this section we solve for the acoustic signal due to arbitrary single particles in its most general form using perturbation theory. We first develop a perturbative approximation scheme for the non-homogeneous strongly damped wave equation Eq.~\eqref{eq:wave-general} based on viscosity. Then, we calculate Green's functions for the retarded propagator on each order, and finally, we provide an explicit solution for the first order in viscosity. 

Throughout the section we will focus on the following form of Eq.~\eqref{eq:wave-source} where the source term has been replaced by an arbitrary function $f : \mathbb{R}^4 \to \mathbb{R}$ such that
\begin{equation}
    \label{eq:wave-general}
    \Box p(\bm{x}) + \lambda \Delta \partial_t p(\bm{x}) = - f(\bm{x}),
\end{equation}
where $\Box$ is the d'Alembert Operator, $\lambda = 1/\omega_0$ is the viscosity coefficient, $p$ is the pressure, and $f$ represents the source and is some function with compact support. Later we will impose more restrictions on $f$ to better represent the energy distribution of a moving particle. However, for now, we consider a general distribution to come up with Green's Functions for the problem.

\subsection{Approximations Using Perturbation Theory\label{subsec:perturbation-theory}}

The solution for the pressure wave $p$ in Eq.~\eqref{eq:wave-general} can be found more easily by treating the damping term $\lambda \Delta \partial_t p(\bm{x})$ as a perturbation for small $\lambda$. Specifically, we can write the solution as a function of $\lambda$ so that
\begin{equation}
    p(\bm{x}) = p_0(\bm{x}) + \lambda p_1(\bm{x}) + \cdots = \sum_{n=0}^{\infty} \lambda^n p_n(\bm{x}).
    \label{eq:perturbed-pressure}
\end{equation}
Using Eq.~\eqref{eq:perturbed-pressure} we can rewrite Eq.~\eqref{eq:wave-general} as
\begin{equation}
    \label{eq:wave-perturbed}
    \Box p_0(\bm{x}) + \sum_{n=1}^{\infty}\lambda^n \left[\Box p_n(\bm{x}) + \Delta \partial_t p_{n-1}(\bm{x}) \right] = -f(\bm{x}).
\end{equation}
For all values of $\lambda$ we can derive the following recursive formula for the $n$th order correction in the pressure
\begin{align}
    \label{eq:wave-perturb-recursive-1}
    \Box p_0(\bm{x}) &= - f(\bm{x})\\
    \label{eq:wave-perturb-recursive-2}
    \Box p_n(\bm{x}) &= - \Delta \partial_t p_{n-1} (\bm{x}).
\end{align}
Using these expressions we can write the following partial differential equation
\begin{equation}
    \label{eq:wave-perturb}
    \Box^{n+1} p_n(\bm{x}) = - \left(- \Delta \partial_t\right)^n f(\bm{x}).
\end{equation}
Note that for $n=0$ we have the normal wave equation with source function $f$. We also see that higher order corrections in pressure are waves that are generated by the higher orders of the curvature of the source function. In other words, if $f$ takes the form of a bump function we will end up adding sharper and more oscillatory corrections for each order.

We will now focus on solving the global Cauchy problem for Eq.~\eqref{eq:wave-perturb}. To do that we will use the method of Green's functions. Before we do so, it is useful to simplify Eq.~\eqref{eq:wave-perturb} using a Fourier transform defined as
\begin{align}
    \label{eq:fourier}
    \hat{f}(\bm{k}) &= \int_{\mathbb{R}^4}d^4x\, f(\bm{x}) e^{-i \bm{k} \cdot \bm{x}}\\
    \label{eq:fourier-inv}
    f(\bm{x}) &= \int_{\mathbb{R}^4}\frac{d^4 k}{(2\pi)^4} f(\bm{k}) e^{i \bm{k} \cdot \bm{x}},
\end{align}
where $\bm{k} \cdot \bm{x}$ is the four-vector inner product defined by $\bm{k} \cdot \bm{x} = k^\alpha x_\alpha = \eta_{\alpha \beta} k_\alpha x_\beta$, where $\eta_{\alpha \beta} = \text{diag}(-1,1,1,1)$ is the Minkowski metric. Using this transformation Eq.~\eqref{eq:wave-perturb} becomes
\begin{equation}
    \label{eq:wave-perturb-ft}
    \hat{p}_n(\bm{k}) = \frac{1}{-\bm{k}^2} \left(\frac{i k_0\,\vec{k}^2}{-\bm{k}^2}\right)^n \hat{f}(\bm{k}),
\end{equation}
where $\vec{k}^2 = k_1^2+  k_2^2 + k_3^2$ and $\bm{k}^2 = -k_0^2 + \vec{k}^2$. This is a much-simplified formula that completely defines the frequency profile of each term given a source function $f$. Now we are ready to solve for the Green's functions.

\subsection{Green's Functions}
\label{subsec:greens-functions}

To provide closed form solutions to Eq.~\eqref{eq:perturbed-pressure} we study a general expression for the retarded propagator of Eq.~\eqref{eq:wave-perturb}. We shall see that the retarded propagator is key to maintaining causality. 

\subsubsection{Using Potentials and Residues}

The form of the propagator greatly simplifies if, instead of solving Eq.~\eqref{eq:wave-perturb} for the $n^{\text{th}}$ pressure component $p_n$ in the perturbative expansion, we solve it for a potential $\psi_{n}$ such that $\partial_t^n \psi_{n} = p_{n}$. With this substitution Eq.~\eqref{eq:wave-perturb} can be rewritten as
\begin{equation}
    \label{eq:wave-perturb-potential}
    \Box^{n+1} \psi_{n}(\bm{x}) = -(-\Delta)^n f(\bm{x}).
\end{equation}
To calculate the Green's function $G_{n}(\bm{x})$ for the $n$th order, we use the Fourier transform of \eqref{eq:fourier}. By plugging in a $\delta$ source, $f(\bm{x}) = \delta (\bm{x})$, in Eq.~\eqref{eq:wave-perturb-potential} and taking the transformation, we obtain an expression for the Fourier transformed Green's function $\hat{G}_n$ 
\begin{equation}
    \label{eq:greens-ft}
    \hat{G}_n(\bm{x}) = \frac{1}{-\bm{k}^2} \left(\frac{\vec{k}^2}{-\bm{k}^2}\right)^n.
\end{equation}
\begin{figure}[!t]
    \centering
    \begin{tikzpicture}[>=latex,scale=1]
\def\R{3}
\def\r{0.1*\R}
\def\k{0.5*\R}
\def\w{1pt}
\def\e{0.03*\R}

\draw [help lines,->](-1.25*\R,0) -- (1.25*\R,0);
\draw [help lines,->](0,-0.4*\R) -- (0,1.25*\R);
\node at (1.25*\R,-0.01*\R)[anchor=north east]{$\textrm{Re }k_0$};
\node at (-0.01*\R,1.25*\R)[anchor=north east]{$\textrm{Im }k_0$};

\path [draw, line width=\w, decoration={
    markings,
    mark=at position 0.450 with {\arrow[line width=\w]{>}},
    mark=at position 0.628 with {\arrow[line width=\w]{>}},
    mark=at position 0.680 with {\arrow[line width=\w]{>}},
    mark=at position 0.760 with {\arrow[line width=\w]{>}},
    mark=at position 0.830 with {\arrow[line width=\w]{>}},
    mark=at position 0.888 with {\arrow[line width=\w]{>}},
    mark=at position 0.970 with {\arrow[line width=\w]{>}},
},postaction=decorate   ]
    (\R,0) arc (0:180:\R) -- (-\R+\k-\r,0) arc (180:0:\r) -- (\R-\k-\r,0) arc (180:0:\r) -- (\R,0);

\node at (-\k,0) {\textbullet};
\node at (-\k,0)[anchor=north] {$-|\vec{k}|$};
\node at (+\k,0) {\textbullet};
\node at (\k,0)[anchor=north] {$|\vec{k}|$};

\draw[->](0,0) -- ({\R * cos(45)},{\R * sin(45)});
\node at ({\R/2 * cos(45)},{\R/2 * sin(45)})[anchor=south east]{$R$};

\draw[-](-\k,0) -- ({-\k+\r * cos(70)},{\r * sin(70)});
\draw[->](-\k+\r+\e,\r/2+1.5*\e) arc (310:290:1);
\node at (-\k+\r+\e,\r/2+1.5*\e)[anchor=south west,shift={(-\e,-\e)}]{\small $\varepsilon$};

\draw[-](\k,0) -- ({\k+\r * cos(70)},{\r * sin(70)});
\draw[->](\k+\r+\e,\r/2+1.5*\e) arc (310:290:1);
\node at (\k+\r+\e,\r/2+1.5*\e)[anchor=south west,shift={(-\e,-\e)}]{\small $\varepsilon$};

\node at ({\R * cos(135)},{\R * sin(135)})[anchor=south east]{$\gamma_{R}$};
\node at (-\k-0.5*\e, 3.5*\e)[anchor=south]{\small $\upmu_-$};
\node at (\k-0.5*\e, 3.5*\e)[anchor=south]{\small $\upmu_+$};

\end{tikzpicture}
    \caption{Integration Contour $\Gamma_\varepsilon$ of $k_0$ in the inverse Fourier transform of Eq.~\ref{eq:greens-ft} to obtain the retarded propagator.}
    \label{fig:retarded-contour}
\end{figure}
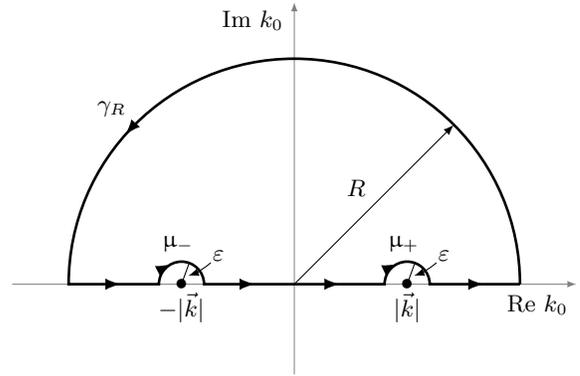
Notice that the singularities of the order $n+1$ are at $\bm{k}^2 = 0$, i.e. for $k_0 = \pm |\vec{k}|$. To extract the retarded propagator in spatial coordinates from $\hat{G}$ we need to take the inverse Fourier transform defined in Eq.~\eqref{eq:fourier-inv} with the contour $\Gamma_\varepsilon$, shown in Fig.~\ref{fig:retarded-contour}, for $\varepsilon \to 0$ to preserve causality. Specifically, such inverse Fourier transform is given by
\begin{equation}
    \label{eq:greens-ift}
    G_n(\bm{x}) = \lim_{\varepsilon \to 0} \int_{\Gamma_\varepsilon} \frac{dk_0}{2\pi}e^{-i k_0 t} \int_{\mathbb{R}^3} \frac{d^3 k}{(2\pi)^3} \hat{G}_n(\bm{k})\, e^{i \vec{k}\cdot \vec{x}}.
\end{equation}
We can rewrite this integral by ``nudging" the singularities by $\varepsilon$. To do so, we do a coordinate transformation by changing $k_0 \to k_0 + i \varepsilon$ so that the integral becomes
\begin{equation}
    \label{eq:greens-ift-2}
    G_n(\bm{x}) = \lim_{\varepsilon \to 0}\int_{\mathbb{R}^4} \frac{d^4 k}{(2\pi)^4} \frac{(\vec{k}^2)^n\, e^{i \bm{k}\cdot \bm{x}}}{\left[(k_0 + i \varepsilon)^2 - \vec{k}^2\right]^{n+1}},
\end{equation}
where the singularities are at $k_0 = \pm |\vec{k}| - i\varepsilon$. We can now do the $k_0$ integral using the residue theorem. Since the analytic form of the residues is quite convoluted, we shall first look at some of their properties as follows.

\subsubsection{Properties of Residues}

As can be seen from Fig.~\ref{fig:retarded-contour}, both singularities are on the lower half of the complex plane. For $t<0$, where the integral converges at the upper half plane, the contour encloses no singularities so that the integral is 0. For $t>0$ the contour includes the singularities and the integral must be calculated. 
Let us denote the two residues by $\hat{g}_\pm(\vec{k}^2)$ corresponding to the singularities at $\pm |\vec{k}| + i \varepsilon$, respectively. Clearly, they are functions of $\vec{k}^2$ as well as $n$ and $\varepsilon$. By using the residue theorem around a circle  small enough to contain only one of the singularities, we can prove that the residues are related as follows
\begin{equation}
    \label{eq:residues-relation}
    \hat{g}_+(\vec{k}^2) = - \hat{g}_-^*(\vec{k}^2),
\end{equation}
where the $^*$ denotes the complex conjugate. This is really convenient as we can rewrite the integral in Eq.~\eqref{eq:greens-ift-2} using the residue theorem as
\begin{equation}
    \label{eq:greens-ift-3}
    G_n(\bm{x}) = \Theta(t)\, \lim_{\varepsilon \to 0}\int_{\mathbb{R}^3} \frac{ d^3 k}{(2\pi)^3} \hat{g}_n(\vec{k}^2) e^{i \vec{k}\cdot \vec{x}},
\end{equation}
where $\Theta(t)$ is the Heaviside function, and $\hat{g}_n(\vec{k}^2)$ is the sum of the residues which is given by
\begin{equation}
    \label{eq:lame-f}
    \hat{g}_n(\vec{k}^2) = i \hat{g}_+ (\vec{k}^2) - i\hat{g}_+^*(\vec{k}^2) = - 2 \text{Im}{\hat{g}_+(\vec{k}^2)},
\end{equation}
where the last step is carried out using $\eqref{eq:residues-relation}$. In this way, not only have we managed to simplify the transform, but we have also shown that the Green's function is real for all orders $n$. That is because $\hat{g}_n$ is radially symmetric in $k$ and real-valued (since it is the imaginary part of a complex-valued function), so the inverse Fourier transform must be real. 

\subsubsection{Calculation of Residues and Green's Functions}

We are now ready to calculate the analytic expression of the residues at the two singularities where $k_0 = k_{\pm} = \pm |\vec{k}| + i \varepsilon$. From what we have shown above, it is enough to calculate the residue at $k_0 = k_+$. We do so using the Laurent expansion of the integrand in Eq.~\eqref{eq:greens-ift-2}. By calculating a series representation centered at $k_0 = k_+$ for each term of the equation, and then multiplying them together, we obtain that the coefficient of the $(k_0 - k_+)$ term is given by
\begin{equation}
    \label{eq:residue}
    \begin{aligned}
    \hat{g}_+ &= \lim_{\varepsilon \to 0} -e^{-\varepsilon t}\Theta(t)\,\text{Im}\, \frac{e^{i (\vec{k}\cdot \vec{x} - |\vec{k}| t)}}{2^n |\vec{k}|} P(|\vec{k}|)\\
    &= - \Theta(t)\,\frac{\sin(\vec{k}\cdot{x} - |\vec{k}| t)}{2^n |\vec{k}|} P(|\vec{k}|),
    \end{aligned}
\end{equation}
where $P(k)$ denotes a polynomial of $|\vec{k}|$ given by
\begin{equation}
    \label{eq:polynomial}
    P(|\vec{k}|) = (-1)^n \sum_{m=0}^{n} \frac{(2i|\vec{k}|t)^m}{m!}\binom{2n - m}{n}.
\end{equation}

We can now plug the expression of the residue into the integral using spherical coordinates to obtain the final expression of the Green's function $G_n$ for the potential $\psi_n$:
\begin{equation}
    \label{eq:retarded-propagator-full}
    \begin{aligned}
    G_n(\bm{x}) = \frac{\Theta(t)}{4^{n+1}\pi r} \sum_{m=0}^{n} \frac{(-2t)^m}{m!} \binom{2n-m}{n}\\ \left[\delta^{(m)}(t-r) - \delta^{(m)}(t+r)\right],
    \end{aligned}
\end{equation}
where $r = |\vec{x}|$. Using $\Theta(t)$ we can simplify the expression above and, by grouping the constants, we obtain
\begin{equation}
    \label{eq:retarded-propagator}
    G_n(\bm{x}) = \sum_{m=0}^n \frac{(-2)^m}{4^{n+1} \pi m!}\binom{2n-m}{n} G_{nm}(\bm{x}),
\end{equation}
where $G_{nm}(\bm{x})$ is given by
\begin{equation}
    \label{eq:reatarded-propagator-terms}
    G_{nm}(\bm{x}) = \frac{t^m}{r} \delta^{(m)}{(t-r)}.
\end{equation}
We managed to express $G_{nm}$ by using the fact that $\Theta(t) \delta(r+t) = 0$ $\forall r>0$. Using the above Green's functions leads to an expression for the pressure
\begin{equation}
    \label{eq:greens-solutions}
    p(\bm{x}) = \sum_{n=0}^{\infty} \partial_{t}^{n}\,(G_n * f)(\bm{x}),
\end{equation}
which is valid for any arbitrary source function $f$.

\subsection{Causality\label{subsec:causality-proof}}

By factoring out a Heaviside theta in Eq.~\eqref{eq:retarded-propagator} it becomes apparent that the propagator is indeed causal. The covolution of Eq.~\eqref{eq:retarded-propagator} with the source function $f$ leads to the following integral
\begin{equation}
    \label{eq:convolution-integral}
    p_n(\bm{x}) = \int_{-\infty}^{t}dt' \int_{\mathbb{R}^3}d^3x'\, G_n(\bm{x} - \bm{x}') f(\bm{x}'),
\end{equation}
from which we can see that each perturbation order in pressure $p_n$ depends on the shape of the source before the time $t$.

\subsection{Closed Form solution for Delta Source}
\label{subsec:closed-form-solution}

Thanks to the analytic expression for the $n$-th order calculated in Eq.~\eqref{eq:retarded-propagator}, we can find a solution for a delta source function for $f$ in closed form. In particular, we want to describe the energy density of the source over space as a moving delta function of the form $\delta(x)\delta(y)\delta(z-v t)$ where $v$ is the speed of the particle in the units where the speed of sound $c$ is equal to 1. Assuming that the energy density that the particle deposits as a function of time is given by a function $q(t)$ we can proceed by defining $f$ as
\begin{equation}
    \label{eq:delta-source}
    f(\bm{x}) = q(t)\delta(x) \delta(y) \delta(z-vt).
\end{equation}

To find $p_n(\bm{x})$ using Eq.~\eqref{eq:convolution-integral}, we need to calculate the convolution $(G_n * f)(\bm{x})$. Since the convolution is a linear operation, we can calculate it by obtaining the convolution for each component $G_{nm}$ in the expansion shown in Eq.~\eqref{eq:retarded-propagator}. Each term of the convolution can be computed by
\begin{equation}
    \label{eq:source-convolution-1}
    G_{nm}*f(\bm{x}) = \int_{\mathbb{R}^3} d^{3}x' \int_{\mathbb{R}}dt'\, G_{nm}(\bm{x}') f(\bm{x}-\bm{x}'),
\end{equation}
where the primed variables in $\bm{x}'$ are the convolution variables. Plugging the functions in, we obtain the following expression
\begin{multline}
    \label{eq:source-convolution-2}
    G_{nm}*f(\bm{x}) = \sum_{l=0}^{m}\sum_{z_k' \in S} (-1)^{l} l! \binom{m}{l}^2 v^{m-l}\\
    \partial_{z'}^{m-l}\left[\frac{q(t-r)r^{m-l}}{r-vz'}\right](z_k'),
\end{multline}
where $v$ is the speed of the particle, $r = \sqrt{x^2 + y^2 + z'^2}$, and $S$ is the set of solutions $z_k'$ of the equation
\begin{equation}
    \label{eq:particle-solutions}
    z' - (z-v t) = v r(z'),
\end{equation}
where $r(z')$ is the radius as a function of $z'$ given by $r(z')~=~\sqrt{x^2 + y^2 + z'^2}$. We can use this expression to more compactly write the convolution $\psi_{n} = G_{n} * f$ so that $\psi_n (\bm{x})$ reads
\begin{equation}
    \label{eq:psi_n-delta-source}
    \psi_n (\bm{x}) = \sum_{m,l}^{n,m}\sum_{z_k' \in S} C_{nml}\, \partial_{z'}^{m-l}\left[\frac{q(t-r)(vr)^{m-l}}{r-vz'}\right](z_k'),
\end{equation}
where $C_{nml}$ is a constant factor given by
\begin{equation}
    \label{eq:multiplier-delta-source}
    C_{nml} := \frac{(-1)^{l+m}\,2^m l!}{4^{n+1} \pi m!} \binom{2n-m}{n}\binom{m}{l}^2.
\end{equation}
Equation \eqref{eq:multiplier-delta-source} shows that the potential is given by derivatives of the same function, evaluated at the roots of Eq.~\eqref{eq:particle-solutions}. The physical meaning of the roots as well as their analytic form is described in the next section.

\subsection{Sound sources for different particle speeds}
\label{subsec:roots-for-different-speeds}

In Eq.~\eqref{eq:psi_n-delta-source} we have shown that the derivative is evaluated at certain ``special'' points that are the roots of Eq.~\eqref{eq:particle-solutions}. Solving this equation, we obtain the following two solutions
\begin{equation}
    \label{eq:z_k-primes}
    z_\pm' = \gamma^2 (z-vt) \pm |v|\sqrt{\gamma^4 (z-vt)^2 + \gamma^2 \rho^2},
\end{equation}
where $\gamma^2 = 1/(1-v^2)$, and $\rho = \sqrt{x^2 + y^2}$ is the distance perpendicular to the path of the particle. However, as clearly shown by Eq.~\eqref{eq:particle-solutions} and Eq.~\eqref{eq:z_k-primes}, these solutions do not always apply. To understand which solution is physically meaningful, we go back to the integral representation of the convolution with the source function in Eq.~\eqref{eq:source-convolution-1}.

The wave generated by a moving particle can be thought of as the sum of the sound waves generated by each point along its motion. The form of those sound waves is given by the retarded propagator we have calculated in Eq.~\eqref{eq:retarded-propagator}. Therefore, when an observer at spacetime point $\bm{x}$ observes a sound wave, we can trace it back to a point in the particle's track where it was generated.

An equivalent way to visualize this is sending particles along the path of a prototype wave. This is the approach we have implemented in the integral in Eq.~\eqref{eq:source-convolution-1}. The equivalence is made rigorous by the commutativity of the convolution. Therefore, Eq.~\eqref{eq:particle-solutions} predicts the points of intersection between a virtual particle passing through an observer located at $\bm{x}$ and a prototype wave propagating from the origin (and described by the propagator). This is clearly shown in Fig.~\ref{fig:convolution}.

\begin{figure}[t!]
    \centering
    \begin{tikzpicture}[>=latex,scale=1]

\def\R{4}
\def\r{0.2*\R}
\def\theta{60}
\def\x{-.15*\R}
\def\y{0.5*\R}
\coordinate (A) at (\R,\R);
\coordinate (B) at (-\R,\R);
\coordinate (a) at ({-\r*cos(\theta) + \x - \y*cos(\theta)},-\r);
\coordinate (b) at ({\R*cos(\theta)  + \x - \y*cos(\theta)},\R);

\coordinate (O) at (\x,\y);

\draw [help lines,->](-\R,0) -- (\R,0);
\draw [help lines,->](0,-\r) -- (0,\R);
\node at (\R,-0.01*\R)[anchor=north east]{$z$};
\node at (-0.01*\R,\R)[anchor=north east]{$t$};

\draw[help lines,dashed,name path=positive] (0,0) -- (A);
\node at (0.9*\R,0.9*\R)[anchor=north,rotate=45] {$z = t$};
\draw[help lines,dashed,name path=negative] (0,0) -- (B);
\node at (-0.9*\R,0.9*\R)[anchor=north,rotate=-45] {$z = -t$};

\draw[thick,name path=obs] ({-\r*cos(\theta)},-\r) -- ({\R*cos(\theta)},\R);
\node at ({0.9*\R*cos(\theta)},0.9*\R)[anchor=north,rotate=\theta] {$z = v t$};

\draw[thick,dotted,name path=particle] (a) -- (b);

\node at (O) {\textbullet};
\node at (O)[anchor=south,shift={(0,0.03*\R)}] {$O$};

\path [name intersections={of=particle and negative,by=E1}];
\node at (E1) {\textbullet};
\node at (E1)[anchor=east,shift={(-0.02*\R,-0.01*\R)}] {$P_+$};

\node at (0,-1.6*\r) {(a) $v<c$};

\end{tikzpicture}
    \vfill
    \begin{tikzpicture}[>=latex,scale=1]

\def\R{4}
\def\s{2}
\def\r{0.2*\R}
\def\theta{10}
\def\y{0.4*\R}
\coordinate (A) at (\R,\R);
\coordinate (B) at (-\R,\R);

\coordinate (a) at (-\R,{-\R*sin(\theta) + \y});
\coordinate (b) at (\R,{\R*sin(\theta) +\y});
\coordinate (c) at (\r - \R,-\r);
\coordinate (d) at (-\R,0);

\coordinate (O) at (-.1*\R,0.95*\y);

\draw [help lines,->](-\R,0) -- (\R,0);
\draw [help lines,->](0,-\r) -- (0,\R);
\node at (\R,-0.01*\R)[anchor=north east]{$z$};
\node at (-0.01*\R,\R)[anchor=north east]{$t$};

\draw[help lines,dashed,name path=positive] (0,0) -- (A);
\node at (0.9*\R,0.9*\R)[anchor=north,rotate=45] {$z = t$};
\draw[help lines,dashed,name path=negative] (0,0) -- (B);
\node at (-0.9*\R,0.9*\R)[anchor=north,rotate=-45] {$z = -t$};

\draw[thick,name path=obs] (-\R,{-\R*sin(\theta)}) -- (\R,{\R*sin(\theta)});
\node at ({0.9*\R*cos(\theta)},{0.9*\R*sin(\theta)})[anchor=north,rotate=\theta] {$z = v t$};

\draw[thick,dotted,name path=particle] (a) -- (b);

\node at (O) {\textbullet};
\node at (O)[anchor=south,shift={(0,0.03*\R)}] {$O$};

\path [name intersections={of=particle and negative,by=E1}];
\node at (E1) {\textbullet};
\node at (E1)[anchor=north,shift={(-0.02*\R,-0.01*\R)}] {$P_-$};

\path [name intersections={of=particle and positive,by=E2}];
\node at (E2) {\textbullet};
\node at (E2)[anchor=north,shift={(0.04*\R,-0.01*\R)}] {$P_+$};

\node at (0,-1.6*\r) {(b) $v>c$};

\end{tikzpicture}
    \caption{Minkowski diagrams for sources moving slower (a) or faster (b) than the speed of sound $c=1$ in the medium. We can see that for slow particles (a) an observer $O$ observes the sound emitted from $P_+$. However, when the source moves faster than the speed of sound (b) the observer at $O$ observes the sound emitted from both $P_-$ and $P_+$. In Eq.~\eqref{eq:particle-solutions} we solve for the z-coordinate of these points.} 
    \label{fig:convolution}
    \bigbreak
    \begin{tikzpicture}[>=latex,scale=1]

\def\R{4}
\def\s{2}
\def\r{0.2*\R}
\def\theta{10}
\coordinate (A) at (\R,\R);
\coordinate (B) at (-\R,\R);

\coordinate (a) at (-\r +\R,-\r);
\coordinate (b) at (\R +\R,\R);
\coordinate (c) at (\r - \R,-\r);
\coordinate (d) at (-\R,0);

\coordinate (O) at (-.1*\R,0.6*\R);

\begin{axis}[
axis x line=center,
axis y line=center,
axis line style ={help lines,->},
xtick=\empty,
ytick=\empty,
xlabel={$z$},
ylabel={$t$},
xlabel style={below},
ylabel style={left},
xmin=-\R,
xmax=\R,
ymin=-\r,
ymax=\R,
unit vector ratio*=1 1 1,
]

\addplot [
    domain=-\R:\R,
    samples=101,
    color=black,
]{abs(x)} node[anchor=north east,shift={(-50,-50)},rotate=45] {$\rho=0$};

\foreach \n in {1, ..., 5}{
    \addplot [
        domain=-\R:\R,
        samples=100,
        color=black,
        style=dashed,
    ]{sqrt(x^2+\n)};
}

\addplot[
    domain=-\R:\R,
    samples=100,
    color=black,
    style=thick,
]{x*tan(\theta)} node[anchor=south east,rotate=\theta] {$z = v t$};

\end{axis}






\end{tikzpicture}
    \caption{The prototype generated wave on the $z-t$ plane as a family of curves for progressively larger perpendicular radii $\rho$ starting from 0. We can see that if $v>1$ there are observers between the solid and dashed lines, but for larger values of $\rho$ there are no intersections with the particle's path.}
    \label{fig:prototype-wave}
\end{figure}

\subsubsection{Slow Particles}

To find out which of the two solutions in Eq.~\eqref{eq:z_k-primes} are applicable in the case of the particle speed being smaller than the speed of sound ($v<1$), we notice the following. Assuming $v<1$ implies that $\gamma^2 > 0$, hence $z_\pm'$ is defined for all values of $z,t,$ and $r$. However, Eq.~\eqref{eq:particle-solutions} imposes another condition which can be written as
\begin{equation}
    \label{eq:z_k-condition}
    z'-(z-vt) > 0.
\end{equation}
It follows that the only solution that satisfies Eq.~\eqref{eq:particle-solutions} is $z_+'$ ($S=\{z_+'\}$ for $v<1$), which is also schematically shown in Fig.~\ref{fig:convolution}.

\subsubsection{Fast Particles}

In the case of particles moving faster than the speed of sound ($v>1$) the considerations become more niche. Fig.~\ref{fig:prototype-wave} illustrates that there are some solutions at positive times that the signal will never reach. Mathematically, this happens when the square root of $z_\pm'$ in Eq.~\eqref{eq:z_k-primes} becomes imaginary. This is possible since $v>1 \implies \gamma^2 <0$. In such case, the delta function in the convolution will have no zeroes and the signal would be zero. This happens when 
\begin{equation}
    \label{eq:v>1-condition}
    vt < z + \rho \sqrt{v^2 -1}.
\end{equation}
When the condition in Eq.~\eqref{eq:v>1-condition} is satisfied, it can be shown that Eq.~\eqref{eq:z_k-condition} implies that, for $(z-vt)>0$, no solution exists and the pressure generated there is 0. However, when $(z-vt) < 0$, both solutions $z_\pm'$ hold and $S~=~\{z_+',z_-'\}$.

\subsection{Leading Order Analytic Solution}
\label{subsec:leading-order-solution}

To make our model a little less abstract we will explicitly provide a closed form solution for $n=0$, i.e. the undamped wave equation. The propagator is given by choosing $n=0$ in Eq.~\eqref{eq:psi_n-delta-source}:
\begin{equation}
    \label{eq:solution-n=0}
    p(\bm{x}) = \sum_{z_k' \in S}\partial_t\, \frac{q(t-r(z_k'))}{4\pi \left(r(z_k') - v z_k'\right)},
\end{equation}
where the set of solutions $S$ has been defined above depending on the value of $v$ and $\bm{x}$. Assuming that the particle starts depositing energy at time $t=0$ we can set $q(t) = \Theta(t)$ so that the solutions from Eq.~\eqref{eq:solution-n=0} simplify further to
\begin{equation}
    \label{eq:solution-n=0-simple}
    p(\bm{x}) = - \frac{ v \bar{z}\, \Theta( - \bar{z} - \rho \sqrt{v^2 -1} )}{\left[\bar{z}^2 + \rho^2 (1-v^2)\right]^{\frac{3}{2}}},
\end{equation}
where $\bar{z} = z - vt$, $\Theta$ is the Heaviside function, and $r = \sqrt{x^2 + y^2 + z^2}$, that is the radius of the observer from the point where the energy deposition starts. Supersonic particles ($v>1$) produce different signals than subsonic ($v<1$). This is a well-known result, and we can use the classical wave equation solution (term for $n=0$) to predict the point and time when the maximum signal occurs as a function of the distance  from the particle track $\rho$ (applying viscous corrections at a second stage).
For the subsonic case, this is at the point when the derivative of the multi-variable function vanishes in all of its components. This condition is given by
\begin{equation}
\label{eq:subsonic-max-pressure}
(z-vt)^2 = \frac{\rho^2}{2\gamma^2}.
\end{equation}
For supersonic particles, even at points away from $\rho = 0$, we have a singularity in the pressure at the zeroth order. This singularity is located within a shell of radius $\mathbb{R}^3$ defined by $\rho^2 + z^2 = t^2$. Hence, given a value of $\rho$, it follows that the maximum will be located at
\begin{equation}
    \label{eq:supersonic-max-pressure}
    \begin{split}
        z^2 &= -\rho^2 \gamma^2\\
        t &= r.
    \end{split}
\end{equation}
It is worth noticing that, in the supersonic case, $v>1\implies\gamma^2<0$. We now exploit these results to examine the pressure around these special points and estimate the signal created by the particles.

\biblio



\section{Application to MIPs}
\label{sec:applications}

Now that we have a perturbative framework for computing the acoustic signal generated by the passage of particles in liquids, we can apply it to real-life situations, such as in solving for the sound wave generated by cosmic-ray muons. Like most relativistic particles, cosmic-ray muons have mean energy loss rates close to the minimum and thus are said to be minimum ionizing particles, or mips~\cite{Amsler2010}. 

We start by deriving an expression for the distribution of the rate of energy deposition of a mip (i.e. the source term in eq.~\eqref{eq:wave-source}). Then, we simulate the evolution of the sound waves to work out peak pressures at various particle speeds in different materials.

\subsection{Energy Deposition Estimation}
\label{subsec:energy-deposition}
To model the effect of a mip going through a fluid, we will ignore any nonlinear effects, such as direct collisions between the particle and the fluid molecules. In that way, we will be able to analytically describe the average energy deposition of the particle. Multiple attempts to describe the energy deposition profile of a single charged particle can be found in literature~\cite{Lahmann2016,Learned1979}. 
\begin{table*}[t!]
\caption{\label{tab:constants} Constants relevant to the calculation for acoustic signals in various fluids. The constants were found in Refs.~\cite{Huber2009,Wagner2002,Lemmon2022,Sharma2013,Groom2021b,Inui2005,Singh2007,Habashi2013,Davis1967,Rao1983,Lemmon2004,Tegeler1999,Span2000,Lemmon2006,Huber2018}}.
\begin{ruledtabular}
\begin{tabular}{ldddddd}
Name &
  \multicolumn{1}{c}{\begin{tabular}[c]{@{}c@{}}Temperature\footnote{All pressures are at $P = 1\ \text{atm}$}\\ $T\ (K)$\end{tabular}} &
  \multicolumn{1}{c}{\begin{tabular}[c]{@{}c@{}}Density\\ $\rho_0\ (kg\,m^{-3})$\end{tabular}} &
  \multicolumn{1}{c}{\begin{tabular}[c]{@{}c@{}}Sound Speed\\ $c\ (m\,s^{-1})$\end{tabular}} &
  \multicolumn{1}{c}{\begin{tabular}[c]{@{}c@{}}Viscosity\\ $\mu\ (\times 10^{-4}\ Pa\, s)$\end{tabular}} &
  \multicolumn{1}{c}{\begin{tabular}[c]{@{}c@{}}Source Term\\ Multiplier\\ $\frac{\beta}{C_p}\ (\times 10^{-6}\, kg\, J^{-1})$\end{tabular}} &
  \multicolumn{1}{c}{\begin{tabular}[c]{@{}c@{}}Mean Excitation \\ Energy\footnote{The scalar multiplier in front of the source term in Eq.~\eqref{eq:wave-source}}\\ $I\ (eV)$\end{tabular}} \\ \hline
     Water & 298.15 &   997.02 & 1496.60 &  8.90030 & 0.06122 &  79.70\\
  Nitrogen &  77.00 &   807.20 &  853.50 &  1.61980 & 2.72482 &  82.00\\
     Argon &  84.00 &  1415.67 &  861.24 &  2.88490 & 3.98389 & 188.00\\
     Xenon & 165.00 &  2942.40 &  643.27 &  5.10420 & 6.66136 & 482.00\\
   Mercury & 298.15 & 13600.00 & 1450.10 & 16.85000 & 1.27989 & 799.97\\
\end{tabular}
\end{ruledtabular}
\end{table*}
While the Bethe-Bloch formula~\cite{Sternheimer1961} can be used to obtain an accurate estimate of the average energy lost by a charged particle in a medium, it does not give the correct estimate of the most probable energy loss because the energy loss distribution per interaction is a highly skewed Landau~\cite{Amsler2010}. 
So instead of the Bethe-Bloch formula, we shall use the most probable value of the Landau distribution as given by~\cite{Groom2021}
\begin{equation}
    \label{eq:bethe-bloch}
    \left. \frac{dE}{dx}\right|_{M} = \xi\left[\ln \frac{2 m_e c_l^{2} \beta^{2} \gamma^{2}}{I}+\ln \frac{\xi}{I}-\beta^{2}\right],
\end{equation}
where $\xi$ is a characteristic energy given in Ref.~\cite{Groom2021}, $I$ is the mean excitation energy of the liquid, $m_e$ is the mass of an electron, $c_l$ is the speed of light, $\gamma$ is the Lorenz factor, and $\beta = v/c_l$ is the speed of the particle relative to the speed of light. The primary assumptions we make are that the particle travels in a straight line through the fluid (along $\hat{\bm{x}}$) and that it is so energetic that its change in speed is negligible while crossing the medium. In more rigorous terms, we assume that
\begin{equation}
    \frac{d}{dt}\frac{dE}{dx} = 0.
\end{equation}
As a result, the rate at which energy is deposited in the medium is given by
\begin{equation}
    \frac{dE}{dt} = \frac{dx}{dt}\left.\frac{dE}{dx}\right|_{M} = v \left.\frac{dE}{dx}\right|_{M}.
    \label{eq:dE-dt}
\end{equation}
We can set up a cylindrical coordinate system around $\hat{\bm{x}}$ where the particle is always at position $(\rho,\phi,x) = (0,0,vt)$ at time t. What we now need to derive is the rate of change of energy density $\epsilon_t(\bm{x},t) = \epsilon_t(\rho,x,t) = dE/dt\,d\Omega$ in order to plug it into the wave equation \eqref{eq:wave-source}.

To do so, consider the energy deposition in some volume $\Omega$. The rate of energy deposition within the volume can be written as (using \eqref{eq:dE-dt})
\begin{equation}
    \frac{dE}{dt} = \int_{\Omega}\, d\Omega\, v \left.\frac{dE}{dx}\right|_{M} G(\bm{x}),
    \label{eq:epsilon-I}
\end{equation}
where $G(\bm{x})$ is the spatial distribution of the energy deposited by the particle. From \eqref{eq:epsilon-I}, we can derive the rate of change of energy density in cylindrical coordinates:
\begin{equation}
    \epsilon_t(\bm{x},t) = \epsilon_t(\rho,\phi,x,t) = \frac{dE}{dt}G(\rho,\phi,x-vt).
\end{equation}

Inspired by Ref.~\cite{Learned1979,Groom2021}, we choose to describe $G$ by a delta distribution. Since such distribution is axially symmetric we can express $\epsilon_t$ as
\begin{equation}
    \epsilon_t(\rho,x,t) = v\left.\frac{dE}{dx}\right|_{M} \delta^2(\rho)\delta(x - v t),
\end{equation}
where $\rho$ is the perpendicular distance from the particle track, $v$ is the speed of the particle, and $x$ is the distance along the track.
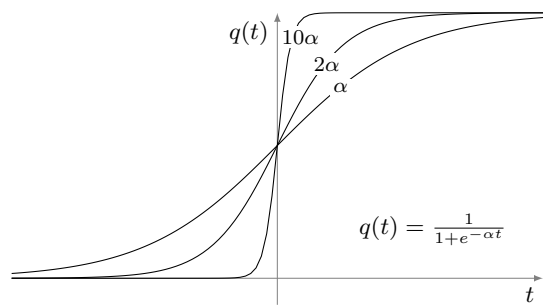
\begin{figure}[b!]
    \centering
    \begin{tikzpicture}[>=latex,scale=1]

\def\T{4}
\def\s{2}
\def\r{0.1*\T}

\begin{axis}[
axis x line=center,
axis y line=center,
axis line style ={help lines,->},
xtick=\empty,
ytick=\empty,
xlabel={$t$},
ylabel={$q(t)$},
xlabel style={below left},
ylabel style={below left},
xmin=-\T,
xmax=\T,
ymin=-0.1,
ymax=1,
unit vector ratio*=1 4 1,
]

\addplot [
    domain=-\T:\T,
    samples=101,
    color=black,
]{1/(1+exp(-x))} node[anchor=north east,shift={(-287,-23)},rotate=0,fill=white,font=\footnotesize,inner sep=0.05cm] {$\alpha$};

\addplot [
    domain=-\T:\T,
    samples=101,
    color=black,
]{1/(1+exp(-2*x))} node[anchor=north east,shift={(-300,-15)},rotate=0,fill=white,font=\footnotesize,inner sep=0.05cm] {$2\alpha$};

\addplot [
    domain=-\T:\T,
    samples=101,
    color=black,
]{1/(1+exp(-10*x))} node[anchor=north east,shift={(-331,-5)},rotate=0,fill=white,font=\footnotesize,inner sep=0.05cm] {$10\alpha$};
\end{axis}

\node at (5.6,1) {$q(t) = \frac{1}{1+e^{-\alpha t}}$};

\end{tikzpicture}
    \caption{Activation function for the source term in Eq.~\eqref{eq:source-term} as a function of parameter $\alpha$. The higher $\alpha$ the more $q(t)$ tends to a Heaviside step.} 
    \label{fig:source-term}
\end{figure}

The last ingredient we need for calculating the rate of energy density deposition in the fluid is an activation term $q(t)$. This is a function $q:\mathbb{R}\to[0,1]$ that only depends on the time and is meant to describe when the particle starts depositing energy in the liquid. In Section~\ref{subsec:leading-order-solution}, we used a Heaviside step function for $q$, however, the discontinuity of the Heaviside leads to a singularity in the pressure that is non-physical. A better model for $q$ is a sigmoid function
\begin{equation}
    q(t) = \frac{1}{1+e^{-\alpha t}},
\end{equation}
where $\alpha$ is a parameter that dictates the gradient of the transition as shown in Fig.~\ref{fig:source-term}. This is a free parameter whose value affects the peak pressure of the signal and has to be fixed depending on the liquid target.

Thus the full source term is
\begin{equation}
    \epsilon_t(\rho,x,t) = v\left.\frac{dE}{dx}\right|_{M} q(t) \delta^2(\rho)\delta(x - v t),
    \label{eq:source-term}
\end{equation}
where $\left.\frac{dE}{dx}\right|_{M}$ is the rate of energy deposition by the particle in the liquid. The source term in Eq.~\ref{eq:source-term} is similar to the source term in Eq.~\ref{eq:delta-source} but multiplied by a constant. We shall use this in the following section to calculate the acoustic signals of muons in various media.

\subsection{Numerical Estimates for Muon Signal in Different Media}
\label{subsec:tables}
In this section, we apply the model just developed to calculate the acoustic signals of relativistic muons passing through water, liquid argon, liquid xenon, liquid nitrogen, and mercury.

To make the calculation of the acoustic signals of single charged particles through different fluids, we built a python package that symbolically evaluates any set of values in Eq.~\eqref{eq:psi_n-delta-source} in parallel and then uses the computer's graphics card to calculate the pressure. The code, installation instructions, and tutorials can be found in Ref.~\cite{Oikonomou2021}.

Using our simulation and the constants in Table~\ref{tab:constants}, we are able to calculate quantitative characteristics for the passage of relativistic muons ($\beta = 0.9$) through the fluids listed above. 

\begin{figure}[!t]
    \centering
    \includegraphics[width=\linewidth]{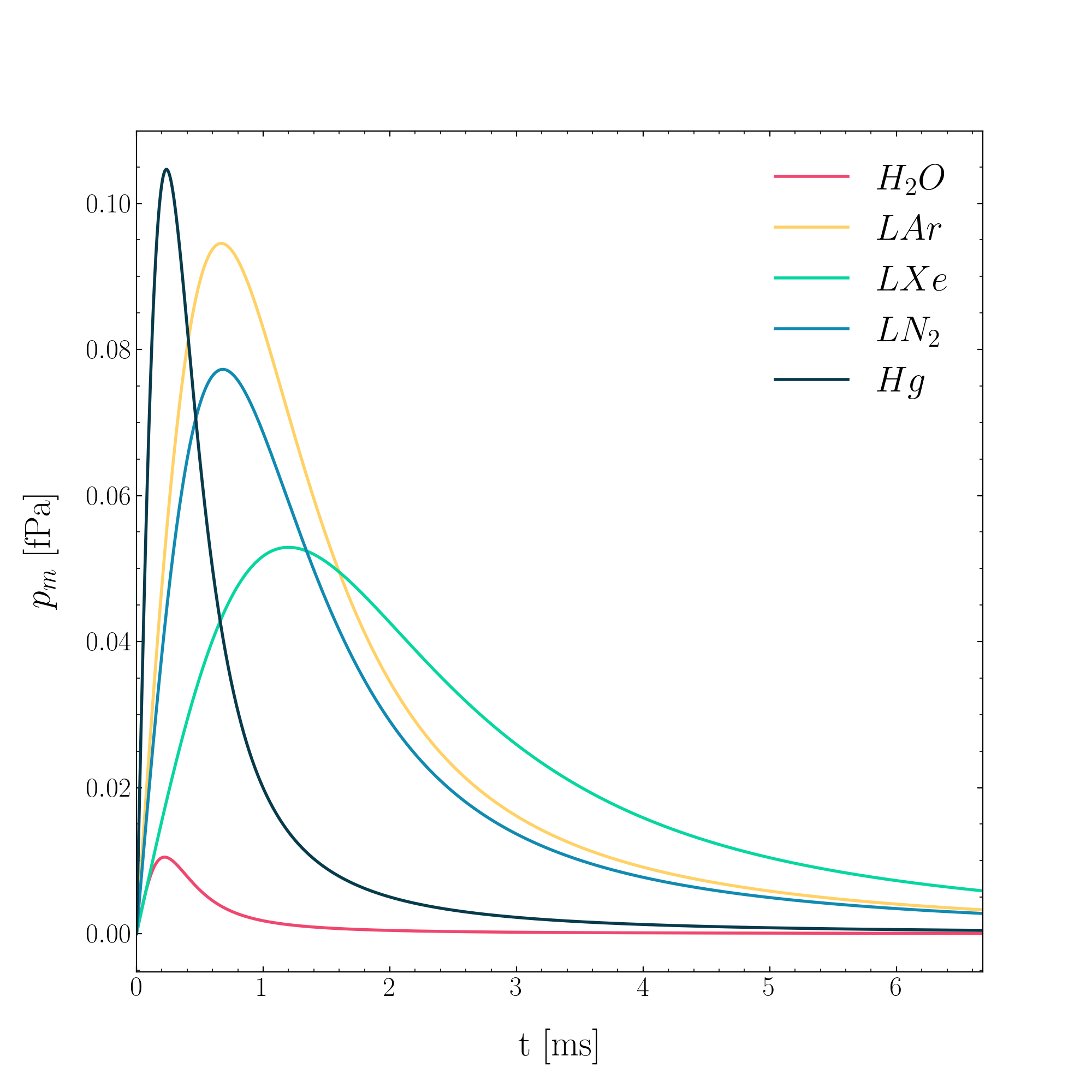}
    \caption{Peak pressure $p_m$ as a function of time $t$ at distance $\rho = 1\ \text{cm}$ in various liquids of the acoustic signal produced by relativistic muons ($\beta = 0.9$). Denser liquids with large coefficient of thermal expansion seem to produce higher peaks for the same particles.} 
    \label{fig:pressure-vs-time}
\end{figure}
\begin{table}[t]
    \centering
    \caption{\label{tab:pressures} Peak pressure $p_m$ of the acoustic signal produced by a single relativistic muon $\beta = 0.9$ in various liquids. The parameters were calculated using the simulation in \cite{Oikonomou2021} and the constants in Table~\ref{tab:constants}.}
    \begin{ruledtabular}
    \begin{tabular}{ldd}
        \multicolumn{1}{c}{\begin{tabular}[c]{@{}c@{}}Name\end{tabular}} &
        \multicolumn{1}{c}{\begin{tabular}[c]{@{}c@{}}Temperature\footnote{All pressures are at $P = 1\ \text{atm}$}\\ $T\ (K)$\end{tabular}} &
        \multicolumn{1}{c}{\begin{tabular}[c]{@{}c@{}}Peak Pressure\\ $p_m\ (\times 10^{-7}\, Pa)$\end{tabular}}\\ \hline
             Water & 298.15 &   1.0499\\
          Nitrogen &  77.00 &   7.7324\\
             Argon &  84.00 &   9.4589\\
             Xenon & 165.00 &   5.2946\\
           Mercury & 298.15 &  10.4703\\
    \end{tabular}
    \end{ruledtabular}
\end{table}
\begin{figure}[!t]
    \centering
    \subfloat[Pressure VS Distance\label{fig:pressure-vs-distance}]{
        \includegraphics[width=\linewidth]{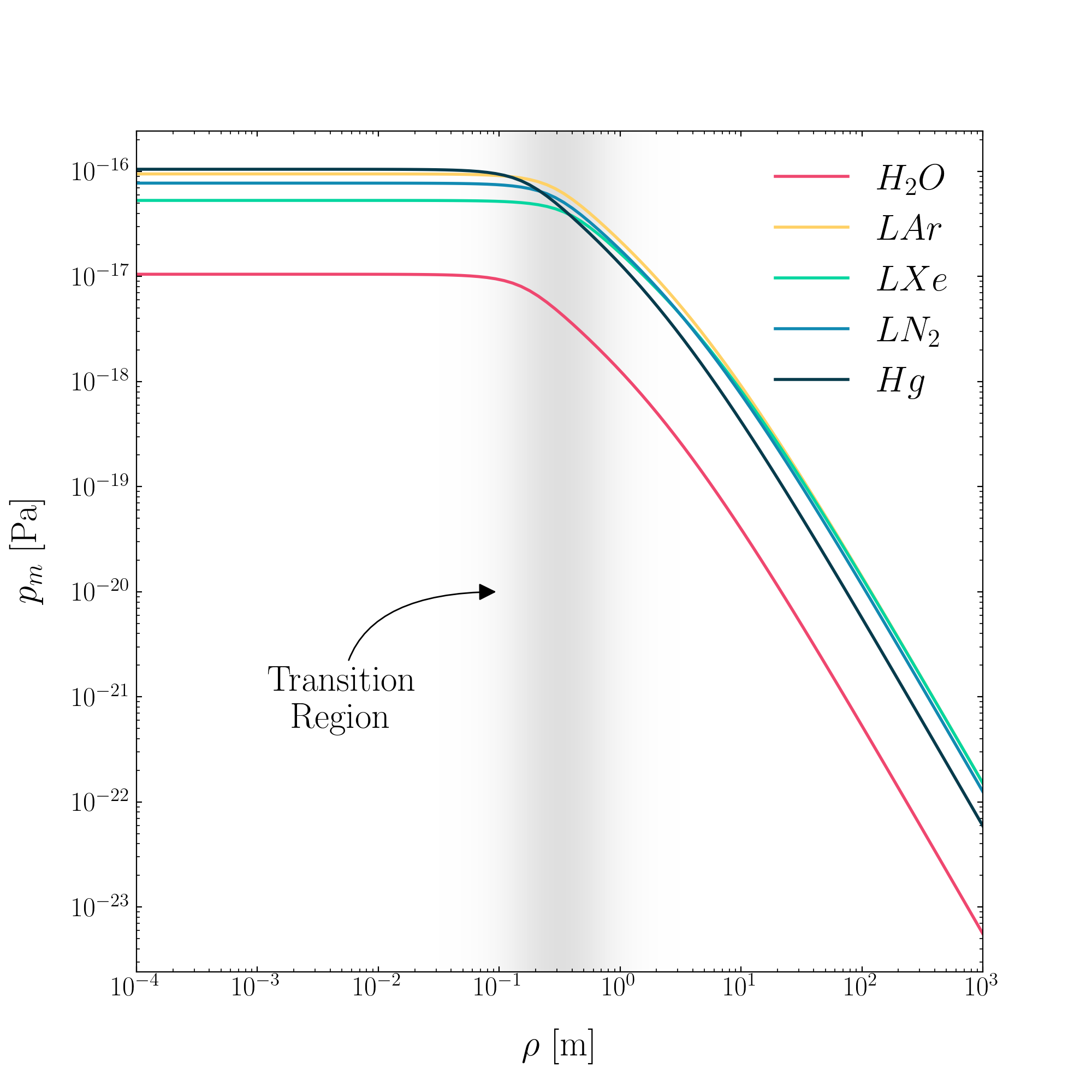}
    }
    \vfill
    \subfloat[Pressure VS Speed\label{fig:pressure-vs-speed}]{
        \includegraphics[width=\linewidth]{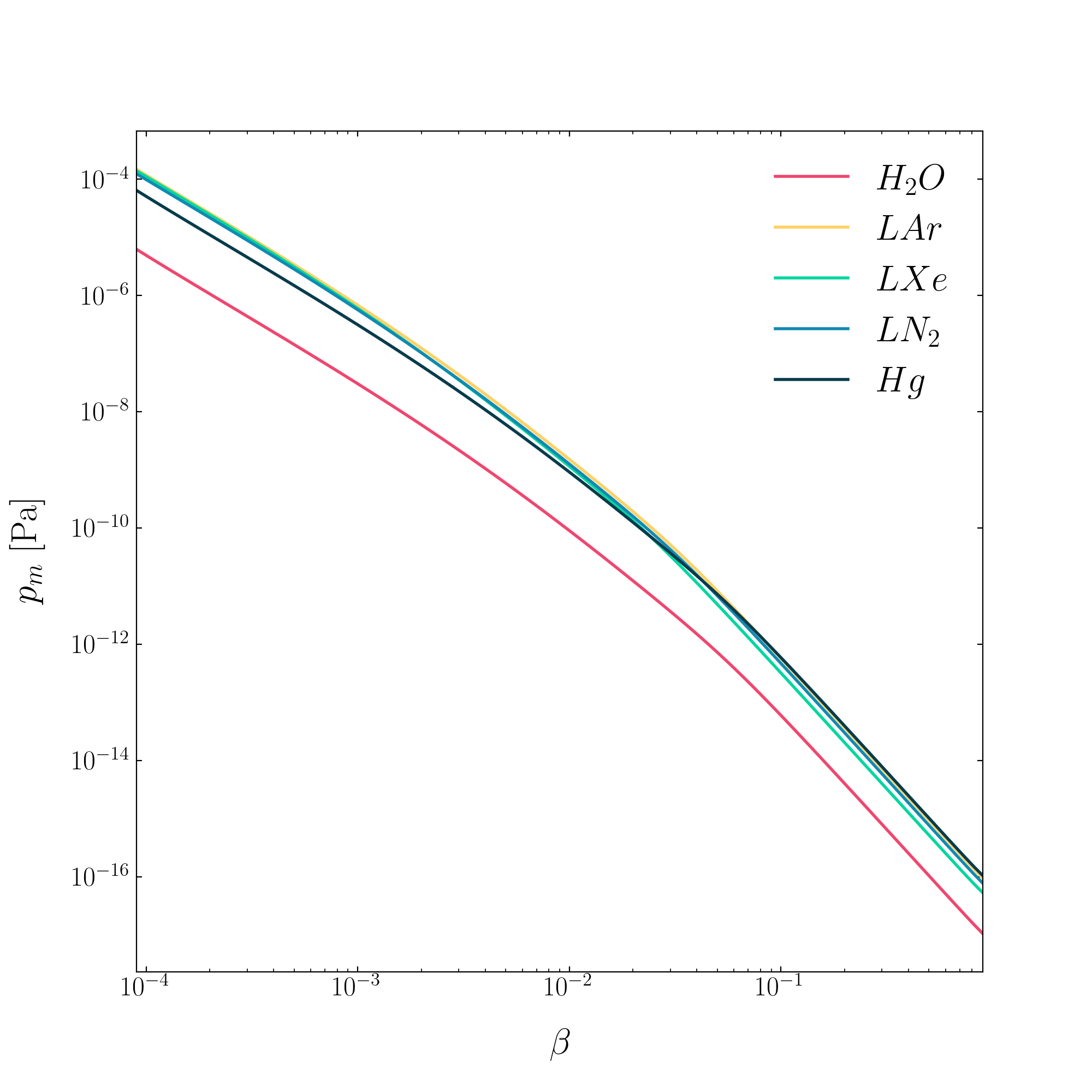}
    }
    \caption{Peak pressure $p_m$ of the acoustic signal produced by muons in various liquids. Subfigure \ref{fig:pressure-vs-distance} presents the peak pressure as a function of distance from the particle track $\rho$ of the acoustic signal produced by relativistic muons ($\beta = 0.9$). Subfigure \ref{fig:pressure-vs-speed} presents the peak pressure as a function of the muon's fractional lightspeed $\beta = v/c_l$.}
    \label{fig:pressure}
\end{figure}

Fig.~\ref{fig:pressure-vs-time} shows the maximum pressure observed $p_m$ at $\rho = 1 \, cm$ away from the particle track as a function of time. The particle generates a skewed pulse over time that peaks at the scale of femto Pascal ($10^-15\, Pa$). Materials with a higher density and coefficient of thermal expansion, like mercury, seem to be the ideal target fluids. Table~\ref{tab:pressures} contains an accurate numerical estimate of the peak pressure at a distance $\rho = 1 cm$ from the track of the muon.

The dependence of the peak pressure on the distance from the particle's track is shown in Fig.~\ref{fig:pressure-vs-distance}. There is a strong exponential trend after $10^{-1}\,m$ while a peak is be reached before that point. The peak pressure signal as a function of velocity of the incoming muons $\beta$, shown in Fig.~\ref{fig:pressure-vs-speed}, is also described by an exponential decay.

\biblio



\section{Conclusion\label{sec:conclusion}}

We have presented a self-contained methodology for calculating acoustic signals from single particle energy depositions in liquids with small, but non-negligible, viscosity. As an example, we have applied our model to predict the signal from relativistic muons crossing multiple liquids. In this analysis, we have only considered the thermal production of sound while ignoring non-linear effects such as microbubble formation. 

The simulation and python package available in Ref.~\cite{Oikonomou2021} can be used to predict the acoustic signal generated by various types of particles in different media. All the assumptions of the model are clearly listed in the derivations provided in the earlier sections, together with a complete description of the physical phenomena at play.

Since we have analytically calculated leading and second order corrections due to viscosity, we can extend the observation made by Learned in Ref.~\cite{Learned1979} to the signals produced by single particles in low-viscous liquids. We have found out that acoustic signals decay with a power law with distance and not of an exponential one as claimed by Ref.~\cite{Learned1979}, making their detection not entirely impossible. However, the signal we predict is of very low frequency and peak amplitude, which is consistent with the very lower energy deposition of single particles.

Further studies are currently being undertaken to completely characterize the effect of viscosity non-perturbatively in order to extend our findings to highly viscous fluids.  Given the small energies and signals involved, we are also developing an Effective Field Theory, that would not only simplify calculations but would also account for quantum effects. From the experimental side, while we acknowledge that this level of signal is too small to be detected with traditional methods such as hydrophones, or sparsely places squids, we believe that other means of detection such as dense sheets of capacitors, superconducting qubits, TES, etc. are worth pursuing. We shall leave the development of the physical sensor to the brave experimental physicists that wish to embark on such journey.

\biblio


\begin{acknowledgments}
We would like to thank Professor John G. Learned (UHM), Dr. Mario Motta (Caltech), and Professor Giorgio Gratta (Stanford University) for the insightful discussions on acoustics. We gratefully acknowledge the support of Professor Elena Beretta (NYUAD), Professor Jonathan Goodman (NYU), Professor Jonathan Weare (NYU), Professor Diogo Arsenio (NYUAD), and Professor Francesco Paparella (NYUAD) in understanding the mathematics behind the physics. Finally, we would like to thank Henry Roberts (NYUAD) and Mariam ElSahhar (NYUAD) for thoroughly proofreading the manuscript, as well as Hashem Al Aidaros (SUAD) for collecting data on liquid constants. This research was carried out on the High Performance Computing resources at New York University Abu Dhabi.
\end{acknowledgments}

\bibliography{reference}

\begin{thebibliography}{46}%
\makeatletter
\providecommand \@ifxundefined [1]{%
 \@ifx{#1\undefined}
}%
\providecommand \@ifnum [1]{%
 \ifnum #1\expandafter \@firstoftwo
 \else \expandafter \@secondoftwo
 \fi
}%
\providecommand \@ifx [1]{%
 \ifx #1\expandafter \@firstoftwo
 \else \expandafter \@secondoftwo
 \fi
}%
\providecommand \natexlab [1]{#1}%
\providecommand \enquote  [1]{``#1''}%
\providecommand \bibnamefont  [1]{#1}%
\providecommand \bibfnamefont [1]{#1}%
\providecommand \citenamefont [1]{#1}%
\providecommand \href@noop [0]{\@secondoftwo}%
\providecommand \href [0]{\begingroup \@sanitize@url \@href}%
\providecommand \@href[1]{\@@startlink{#1}\@@href}%
\providecommand \@@href[1]{\endgroup#1\@@endlink}%
\providecommand \@sanitize@url [0]{\catcode `\\12\catcode `\$12\catcode
  `\&12\catcode `\#12\catcode `\^12\catcode `\_12\catcode `\%12\relax}%
\providecommand \@@startlink[1]{}%
\providecommand \@@endlink[0]{}%
\providecommand \url  [0]{\begingroup\@sanitize@url \@url }%
\providecommand \@url [1]{\endgroup\@href {#1}{\urlprefix }}%
\providecommand \urlprefix  [0]{URL }%
\providecommand \Eprint [0]{\href }%
\providecommand \doibase [0]{https://doi.org/}%
\providecommand \selectlanguage [0]{\@gobble}%
\providecommand \bibinfo  [0]{\@secondoftwo}%
\providecommand \bibfield  [0]{\@secondoftwo}%
\providecommand \translation [1]{[#1]}%
\providecommand \BibitemOpen [0]{}%
\providecommand \bibitemStop [0]{}%
\providecommand \bibitemNoStop [0]{.\EOS\space}%
\providecommand \EOS [0]{\spacefactor3000\relax}%
\providecommand \BibitemShut  [1]{\csname bibitem#1\endcsname}%
\let\auto@bib@innerbib\@empty
\bibitem [{\citenamefont {Askaryan}(1957)}]{Askaryan1957}%
  \BibitemOpen
  \bibfield  {author} {\bibinfo {author} {\bibfnamefont {G.~A.}\ \bibnamefont
  {Askaryan}},\ }\href {https://doi.org/10.1007/BF01480076} {\bibfield
  {journal} {\bibinfo  {journal} {The Soviet Journal of Atomic Energy}\
  }\textbf {\bibinfo {volume} {3}},\ \bibinfo {pages} {921} (\bibinfo {year}
  {1957})}\BibitemShut {NoStop}%
\bibitem [{\citenamefont {Lahmann}\ \emph {et~al.}(2015)\citenamefont
  {Lahmann}, \citenamefont {Anton}, \citenamefont {Graf}, \citenamefont
  {Hößl}, \citenamefont {Kappes}, \citenamefont {Katz}, \citenamefont
  {Mecke},\ and\ \citenamefont {Schwemmer}}]{Lahman2015}%
  \BibitemOpen
  \bibfield  {author} {\bibinfo {author} {\bibfnamefont {R.}~\bibnamefont
  {Lahmann}}, \bibinfo {author} {\bibfnamefont {G.}~\bibnamefont {Anton}},
  \bibinfo {author} {\bibfnamefont {K.}~\bibnamefont {Graf}}, \bibinfo {author}
  {\bibfnamefont {J.}~\bibnamefont {Hößl}}, \bibinfo {author} {\bibfnamefont
  {A.}~\bibnamefont {Kappes}}, \bibinfo {author} {\bibfnamefont
  {U.}~\bibnamefont {Katz}}, \bibinfo {author} {\bibfnamefont {K.}~\bibnamefont
  {Mecke}},\ and\ \bibinfo {author} {\bibfnamefont {S.}~\bibnamefont
  {Schwemmer}},\ }\href
  {https://doi.org/https://doi.org/10.1016/j.astropartphys.2014.12.003}
  {\bibfield  {journal} {\bibinfo  {journal} {Astroparticle Physics}\ }\textbf
  {\bibinfo {volume} {65}},\ \bibinfo {pages} {69} (\bibinfo {year}
  {2015})}\BibitemShut {NoStop}%
\bibitem [{\citenamefont {Roberts}(1992)}]{Roberts1992}%
  \BibitemOpen
  \bibfield  {author} {\bibinfo {author} {\bibfnamefont {A.}~\bibnamefont
  {Roberts}},\ }\href {https://doi.org/10.1103/RevModPhys.64.259} {\bibfield
  {journal} {\bibinfo  {journal} {Rev. Mod. Phys.}\ }\textbf {\bibinfo {volume}
  {64}},\ \bibinfo {pages} {259} (\bibinfo {year} {1992})}\BibitemShut
  {NoStop}%
\bibitem [{\citenamefont {Karle}\ \emph {et~al.}(2003)\citenamefont {Karle},
  \citenamefont {Ahrens}, \citenamefont {Bahcall}, \citenamefont {Bai},
  \citenamefont {Becka}, \citenamefont {Becker}, \citenamefont {Besson},
  \citenamefont {Berley}, \citenamefont {Bernardini}, \citenamefont {Bertrand},
  \citenamefont {Binon}, \citenamefont {Biron}, \citenamefont {Böser},
  \citenamefont {Bohm}, \citenamefont {Botner}, \citenamefont {Bouhali},
  \citenamefont {Burgess}, \citenamefont {Castermans}, \citenamefont {Chirkin},
  \citenamefont {Conrad}, \citenamefont {Cooley}, \citenamefont {Cowen},
  \citenamefont {Davour}, \citenamefont {{De Clercq}}, \citenamefont {DeYoung},
  \citenamefont {Desiati}, \citenamefont {Dewulf}, \citenamefont {Dingus},
  \citenamefont {Ellsworth}, \citenamefont {Evenson}, \citenamefont {Fazely},
  \citenamefont {Feser}, \citenamefont {Gaisser}, \citenamefont {Gallagher},
  \citenamefont {Ganugapati}, \citenamefont {Goldschmidt}, \citenamefont
  {Goodman}, \citenamefont {Hallgren}, \citenamefont {Halzen}, \citenamefont
  {Hanson}, \citenamefont {Hardtke}, \citenamefont {Hauschildt}, \citenamefont
  {Hellwig}, \citenamefont {Herquet}, \citenamefont {Hill}, \citenamefont
  {Hulth}, \citenamefont {Hultgvist}, \citenamefont {Hundertmark},
  \citenamefont {Jacobsen}, \citenamefont {Japaridze}, \citenamefont {Karle},
  \citenamefont {Köpke}, \citenamefont {Kowalski}, \citenamefont {Lamoureux},
  \citenamefont {Leich}, \citenamefont {Leuthold}, \citenamefont {Lindahl},
  \citenamefont {Liubarsky}, \citenamefont {Madson}, \citenamefont
  {Marciniewski}, \citenamefont {Matis}, \citenamefont {McParland},
  \citenamefont {Minaeva}, \citenamefont {Miočinović}, \citenamefont {Morse},
  \citenamefont {Nahnhauer}, \citenamefont {Neunhöffer}, \citenamefont
  {Niessen}, \citenamefont {Nygren}, \citenamefont {Ogelman}, \citenamefont
  {Olbrechts}, \citenamefont {{Pérez de los Heros}}, \citenamefont {Pohl},
  \citenamefont {Price}, \citenamefont {Przybylski}, \citenamefont {Rawlins},
  \citenamefont {Resconi}, \citenamefont {Rhode}, \citenamefont {Ribordy},
  \citenamefont {Richter}, \citenamefont {Sander}, \citenamefont {Schmidt},
  \citenamefont {Schneider}, \citenamefont {Seckel}, \citenamefont {Solarz},
  \citenamefont {Sparke}, \citenamefont {Spiczak}, \citenamefont {Spiering},
  \citenamefont {Stanev}, \citenamefont {Steele}, \citenamefont {Steffen},
  \citenamefont {Stokstad}, \citenamefont {Sudhoff}, \citenamefont {Sulanke},
  \citenamefont {Sullivan}, \citenamefont {Sumners}, \citenamefont {Taboada},
  \citenamefont {Thollander}, \citenamefont {Tilav}, \citenamefont {Walck},
  \citenamefont {Weinheimer}, \citenamefont {Wiebusch}, \citenamefont
  {Wiedemann}, \citenamefont {Wischnewski}, \citenamefont {Wissing},
  \citenamefont {Woschnagg},\ and\ \citenamefont {Yoshida}}]{Karle2003}%
  \BibitemOpen
  \bibfield  {author} {\bibinfo {author} {\bibfnamefont {A.}~\bibnamefont
  {Karle}}, \bibinfo {author} {\bibfnamefont {J.}~\bibnamefont {Ahrens}},
  \bibinfo {author} {\bibfnamefont {J.}~\bibnamefont {Bahcall}}, \bibinfo
  {author} {\bibfnamefont {X.}~\bibnamefont {Bai}}, \bibinfo {author}
  {\bibfnamefont {T.}~\bibnamefont {Becka}}, \bibinfo {author} {\bibfnamefont
  {K.-H.}\ \bibnamefont {Becker}}, \bibinfo {author} {\bibfnamefont
  {D.}~\bibnamefont {Besson}}, \bibinfo {author} {\bibfnamefont
  {D.}~\bibnamefont {Berley}}, \bibinfo {author} {\bibfnamefont
  {E.}~\bibnamefont {Bernardini}}, \bibinfo {author} {\bibfnamefont
  {D.}~\bibnamefont {Bertrand}}, \bibinfo {author} {\bibfnamefont
  {F.}~\bibnamefont {Binon}}, \bibinfo {author} {\bibfnamefont
  {A.}~\bibnamefont {Biron}}, \bibinfo {author} {\bibfnamefont
  {S.}~\bibnamefont {Böser}}, \bibinfo {author} {\bibfnamefont
  {C.}~\bibnamefont {Bohm}}, \bibinfo {author} {\bibfnamefont {O.}~\bibnamefont
  {Botner}}, \bibinfo {author} {\bibfnamefont {O.}~\bibnamefont {Bouhali}},
  \bibinfo {author} {\bibfnamefont {T.}~\bibnamefont {Burgess}}, \bibinfo
  {author} {\bibfnamefont {T.}~\bibnamefont {Castermans}}, \bibinfo {author}
  {\bibfnamefont {D.}~\bibnamefont {Chirkin}}, \bibinfo {author} {\bibfnamefont
  {J.}~\bibnamefont {Conrad}}, \bibinfo {author} {\bibfnamefont
  {J.}~\bibnamefont {Cooley}}, \bibinfo {author} {\bibfnamefont
  {D.}~\bibnamefont {Cowen}}, \bibinfo {author} {\bibfnamefont
  {A.}~\bibnamefont {Davour}}, \bibinfo {author} {\bibfnamefont
  {C.}~\bibnamefont {{De Clercq}}}, \bibinfo {author} {\bibfnamefont
  {T.}~\bibnamefont {DeYoung}}, \bibinfo {author} {\bibfnamefont
  {P.}~\bibnamefont {Desiati}}, \bibinfo {author} {\bibfnamefont {J.-P.}\
  \bibnamefont {Dewulf}}, \bibinfo {author} {\bibfnamefont {B.}~\bibnamefont
  {Dingus}}, \bibinfo {author} {\bibfnamefont {R.}~\bibnamefont {Ellsworth}},
  \bibinfo {author} {\bibfnamefont {P.}~\bibnamefont {Evenson}}, \bibinfo
  {author} {\bibfnamefont {A.}~\bibnamefont {Fazely}}, \bibinfo {author}
  {\bibfnamefont {T.}~\bibnamefont {Feser}}, \bibinfo {author} {\bibfnamefont
  {T.}~\bibnamefont {Gaisser}}, \bibinfo {author} {\bibfnamefont
  {J.}~\bibnamefont {Gallagher}}, \bibinfo {author} {\bibfnamefont
  {R.}~\bibnamefont {Ganugapati}}, \bibinfo {author} {\bibfnamefont
  {A.}~\bibnamefont {Goldschmidt}}, \bibinfo {author} {\bibfnamefont
  {J.}~\bibnamefont {Goodman}}, \bibinfo {author} {\bibfnamefont
  {A.}~\bibnamefont {Hallgren}}, \bibinfo {author} {\bibfnamefont
  {F.}~\bibnamefont {Halzen}}, \bibinfo {author} {\bibfnamefont
  {K.}~\bibnamefont {Hanson}}, \bibinfo {author} {\bibfnamefont
  {R.}~\bibnamefont {Hardtke}}, \bibinfo {author} {\bibfnamefont
  {T.}~\bibnamefont {Hauschildt}}, \bibinfo {author} {\bibfnamefont
  {M.}~\bibnamefont {Hellwig}}, \bibinfo {author} {\bibfnamefont
  {P.}~\bibnamefont {Herquet}}, \bibinfo {author} {\bibfnamefont
  {G.}~\bibnamefont {Hill}}, \bibinfo {author} {\bibfnamefont {P.}~\bibnamefont
  {Hulth}}, \bibinfo {author} {\bibfnamefont {K.}~\bibnamefont {Hultgvist}},
  \bibinfo {author} {\bibfnamefont {S.}~\bibnamefont {Hundertmark}}, \bibinfo
  {author} {\bibfnamefont {J.}~\bibnamefont {Jacobsen}}, \bibinfo {author}
  {\bibfnamefont {G.}~\bibnamefont {Japaridze}}, \bibinfo {author}
  {\bibfnamefont {A.}~\bibnamefont {Karle}}, \bibinfo {author} {\bibfnamefont
  {L.}~\bibnamefont {Köpke}}, \bibinfo {author} {\bibfnamefont
  {M.}~\bibnamefont {Kowalski}}, \bibinfo {author} {\bibfnamefont
  {J.}~\bibnamefont {Lamoureux}}, \bibinfo {author} {\bibfnamefont
  {H.}~\bibnamefont {Leich}}, \bibinfo {author} {\bibfnamefont
  {M.}~\bibnamefont {Leuthold}}, \bibinfo {author} {\bibfnamefont
  {P.}~\bibnamefont {Lindahl}}, \bibinfo {author} {\bibfnamefont
  {I.}~\bibnamefont {Liubarsky}}, \bibinfo {author} {\bibfnamefont
  {J.}~\bibnamefont {Madson}}, \bibinfo {author} {\bibfnamefont
  {P.}~\bibnamefont {Marciniewski}}, \bibinfo {author} {\bibfnamefont
  {H.}~\bibnamefont {Matis}}, \bibinfo {author} {\bibfnamefont
  {C.}~\bibnamefont {McParland}}, \bibinfo {author} {\bibfnamefont
  {Y.}~\bibnamefont {Minaeva}}, \bibinfo {author} {\bibfnamefont
  {P.}~\bibnamefont {Miočinović}}, \bibinfo {author} {\bibfnamefont
  {R.}~\bibnamefont {Morse}}, \bibinfo {author} {\bibfnamefont
  {R.}~\bibnamefont {Nahnhauer}}, \bibinfo {author} {\bibfnamefont
  {T.}~\bibnamefont {Neunhöffer}}, \bibinfo {author} {\bibfnamefont
  {P.}~\bibnamefont {Niessen}}, \bibinfo {author} {\bibfnamefont
  {D.}~\bibnamefont {Nygren}}, \bibinfo {author} {\bibfnamefont
  {H.}~\bibnamefont {Ogelman}}, \bibinfo {author} {\bibfnamefont
  {P.}~\bibnamefont {Olbrechts}}, \bibinfo {author} {\bibfnamefont
  {C.}~\bibnamefont {{Pérez de los Heros}}}, \bibinfo {author} {\bibfnamefont
  {A.}~\bibnamefont {Pohl}}, \bibinfo {author} {\bibfnamefont {P.}~\bibnamefont
  {Price}}, \bibinfo {author} {\bibfnamefont {G.}~\bibnamefont {Przybylski}},
  \bibinfo {author} {\bibfnamefont {K.}~\bibnamefont {Rawlins}}, \bibinfo
  {author} {\bibfnamefont {E.}~\bibnamefont {Resconi}}, \bibinfo {author}
  {\bibfnamefont {W.}~\bibnamefont {Rhode}}, \bibinfo {author} {\bibfnamefont
  {M.}~\bibnamefont {Ribordy}}, \bibinfo {author} {\bibfnamefont
  {S.}~\bibnamefont {Richter}}, \bibinfo {author} {\bibfnamefont {H.-G.}\
  \bibnamefont {Sander}}, \bibinfo {author} {\bibfnamefont {T.}~\bibnamefont
  {Schmidt}}, \bibinfo {author} {\bibfnamefont {D.}~\bibnamefont {Schneider}},
  \bibinfo {author} {\bibfnamefont {D.}~\bibnamefont {Seckel}}, \bibinfo
  {author} {\bibfnamefont {M.}~\bibnamefont {Solarz}}, \bibinfo {author}
  {\bibfnamefont {L.}~\bibnamefont {Sparke}}, \bibinfo {author} {\bibfnamefont
  {G.}~\bibnamefont {Spiczak}}, \bibinfo {author} {\bibfnamefont
  {C.}~\bibnamefont {Spiering}}, \bibinfo {author} {\bibfnamefont
  {T.}~\bibnamefont {Stanev}}, \bibinfo {author} {\bibfnamefont
  {D.}~\bibnamefont {Steele}}, \bibinfo {author} {\bibfnamefont
  {P.}~\bibnamefont {Steffen}}, \bibinfo {author} {\bibfnamefont
  {R.}~\bibnamefont {Stokstad}}, \bibinfo {author} {\bibfnamefont
  {P.}~\bibnamefont {Sudhoff}}, \bibinfo {author} {\bibfnamefont {K.-H.}\
  \bibnamefont {Sulanke}}, \bibinfo {author} {\bibfnamefont {G.}~\bibnamefont
  {Sullivan}}, \bibinfo {author} {\bibfnamefont {T.}~\bibnamefont {Sumners}},
  \bibinfo {author} {\bibfnamefont {I.}~\bibnamefont {Taboada}}, \bibinfo
  {author} {\bibfnamefont {L.}~\bibnamefont {Thollander}}, \bibinfo {author}
  {\bibfnamefont {S.}~\bibnamefont {Tilav}}, \bibinfo {author} {\bibfnamefont
  {C.}~\bibnamefont {Walck}}, \bibinfo {author} {\bibfnamefont
  {C.}~\bibnamefont {Weinheimer}}, \bibinfo {author} {\bibfnamefont
  {C.}~\bibnamefont {Wiebusch}}, \bibinfo {author} {\bibfnamefont
  {C.}~\bibnamefont {Wiedemann}}, \bibinfo {author} {\bibfnamefont
  {R.}~\bibnamefont {Wischnewski}}, \bibinfo {author} {\bibfnamefont
  {H.}~\bibnamefont {Wissing}}, \bibinfo {author} {\bibfnamefont
  {K.}~\bibnamefont {Woschnagg}},\ and\ \bibinfo {author} {\bibfnamefont
  {S.}~\bibnamefont {Yoshida}},\ }\href
  {https://doi.org/https://doi.org/10.1016/S0920-5632(03)01337-9} {\bibfield
  {journal} {\bibinfo  {journal} {Nuclear Physics B - Proceedings Supplements}\
  }\textbf {\bibinfo {volume} {118}},\ \bibinfo {pages} {388} (\bibinfo {year}
  {2003})},\ \bibinfo {note} {proceedings of the XXth International Conference
  on Neutrino Physics and Astrophysics}\BibitemShut {NoStop}%
\bibitem [{\citenamefont {Askaryan}\ \emph {et~al.}(1979)\citenamefont
  {Askaryan}, \citenamefont {Dolgoshein}, \citenamefont {Kalinovsky},\ and\
  \citenamefont {Mokhov}}]{Askaryan1979}%
  \BibitemOpen
  \bibfield  {author} {\bibinfo {author} {\bibfnamefont {G.}~\bibnamefont
  {Askaryan}}, \bibinfo {author} {\bibfnamefont {B.}~\bibnamefont
  {Dolgoshein}}, \bibinfo {author} {\bibfnamefont {A.}~\bibnamefont
  {Kalinovsky}},\ and\ \bibinfo {author} {\bibfnamefont {N.}~\bibnamefont
  {Mokhov}},\ }\href
  {https://doi.org/https://doi.org/10.1016/0029-554X(79)90244-1} {\bibfield
  {journal} {\bibinfo  {journal} {Nuclear Instruments and Methods}\ }\textbf
  {\bibinfo {volume} {164}},\ \bibinfo {pages} {267} (\bibinfo {year}
  {1979})}\BibitemShut {NoStop}%
\bibitem [{\citenamefont {Sulak}\ \emph {et~al.}(1979)\citenamefont {Sulak},
  \citenamefont {Armstrong}, \citenamefont {Baranger}, \citenamefont {Bregman},
  \citenamefont {Levi}, \citenamefont {Mael}, \citenamefont {Strait},
  \citenamefont {Bowen}, \citenamefont {Pifer}, \citenamefont {Polakos},
  \citenamefont {Bradner}, \citenamefont {Parvulescu}, \citenamefont {Jones},\
  and\ \citenamefont {Learned}}]{Sulak1979}%
  \BibitemOpen
  \bibfield  {author} {\bibinfo {author} {\bibfnamefont {L.}~\bibnamefont
  {Sulak}}, \bibinfo {author} {\bibfnamefont {T.}~\bibnamefont {Armstrong}},
  \bibinfo {author} {\bibfnamefont {H.}~\bibnamefont {Baranger}}, \bibinfo
  {author} {\bibfnamefont {M.}~\bibnamefont {Bregman}}, \bibinfo {author}
  {\bibfnamefont {M.}~\bibnamefont {Levi}}, \bibinfo {author} {\bibfnamefont
  {D.}~\bibnamefont {Mael}}, \bibinfo {author} {\bibfnamefont {J.}~\bibnamefont
  {Strait}}, \bibinfo {author} {\bibfnamefont {T.}~\bibnamefont {Bowen}},
  \bibinfo {author} {\bibfnamefont {A.}~\bibnamefont {Pifer}}, \bibinfo
  {author} {\bibfnamefont {P.}~\bibnamefont {Polakos}}, \bibinfo {author}
  {\bibfnamefont {H.}~\bibnamefont {Bradner}}, \bibinfo {author} {\bibfnamefont
  {A.}~\bibnamefont {Parvulescu}}, \bibinfo {author} {\bibfnamefont
  {W.}~\bibnamefont {Jones}},\ and\ \bibinfo {author} {\bibfnamefont
  {J.}~\bibnamefont {Learned}},\ }\href
  {https://doi.org/https://doi.org/10.1016/0029-554X(79)90386-0} {\bibfield
  {journal} {\bibinfo  {journal} {Nuclear Instruments and Methods}\ }\textbf
  {\bibinfo {volume} {161}},\ \bibinfo {pages} {203} (\bibinfo {year}
  {1979})}\BibitemShut {NoStop}%
\bibitem [{\citenamefont {{Jiang}}\ \emph {et~al.}(1985)\citenamefont
  {{Jiang}}, \citenamefont {{Yuan}}, \citenamefont {{Li}}, \citenamefont
  {{Chen}}, \citenamefont {{Zheng}}, \citenamefont {{Song}},\ and\
  \citenamefont {{Jiang}}}]{Jiang1985}%
  \BibitemOpen
  \bibfield  {author} {\bibinfo {author} {\bibfnamefont {Y.~L.}\ \bibnamefont
  {{Jiang}}}, \bibinfo {author} {\bibfnamefont {Y.~K.}\ \bibnamefont {{Yuan}}},
  \bibinfo {author} {\bibfnamefont {Y.~G.}\ \bibnamefont {{Li}}}, \bibinfo
  {author} {\bibfnamefont {D.~B.}\ \bibnamefont {{Chen}}}, \bibinfo {author}
  {\bibfnamefont {R.~T.}\ \bibnamefont {{Zheng}}}, \bibinfo {author}
  {\bibfnamefont {J.~N.}\ \bibnamefont {{Song}}},\ and\ \bibinfo {author}
  {\bibfnamefont {Y.~L.}\ \bibnamefont {{Jiang}}},\ }in\ \href@noop {} {\emph
  {\bibinfo {booktitle} {19th International Cosmic Ray Conference (ICRC19),
  Volume 8}}},\ \bibinfo {series} {International Cosmic Ray Conference},
  Vol.~\bibinfo {volume} {8}\ (\bibinfo {year} {1985})\ p.\ \bibinfo {pages}
  {329}\BibitemShut {NoStop}%
\bibitem [{\citenamefont {{Bowen}}(1979)}]{Bowen1979}%
  \BibitemOpen
  \bibfield  {author} {\bibinfo {author} {\bibfnamefont {T.}~\bibnamefont
  {{Bowen}}},\ }in\ \href@noop {} {\emph {\bibinfo {booktitle} {International
  Cosmic Ray Conference}}},\ \bibinfo {series} {International Cosmic Ray
  Conference}, Vol.~\bibinfo {volume} {11}\ (\bibinfo {year} {1979})\ p.\
  \bibinfo {pages} {184}\BibitemShut {NoStop}%
\bibitem [{\citenamefont {Albul}\ \emph {et~al.}(2001)\citenamefont {Albul},
  \citenamefont {Bychkov}, \citenamefont {Gusev}, \citenamefont {Demidov},
  \citenamefont {Demidova}, \citenamefont {Konovalov}, \citenamefont
  {Kurchanov}, \citenamefont {Luk'yashin}, \citenamefont {Lyashuk},
  \citenamefont {Novikov}, \citenamefont {Rostovtsev}, \citenamefont {Sokolov},
  \citenamefont {Feizkhanov},\ and\ \citenamefont {Khaldeeva}}]{Albul2001}%
  \BibitemOpen
  \bibfield  {author} {\bibinfo {author} {\bibfnamefont {V.}~\bibnamefont
  {Albul}}, \bibinfo {author} {\bibfnamefont {V.}~\bibnamefont {Bychkov}},
  \bibinfo {author} {\bibfnamefont {K.}~\bibnamefont {Gusev}}, \bibinfo
  {author} {\bibfnamefont {V.}~\bibnamefont {Demidov}}, \bibinfo {author}
  {\bibfnamefont {E.}~\bibnamefont {Demidova}}, \bibinfo {author}
  {\bibfnamefont {S.}~\bibnamefont {Konovalov}}, \bibinfo {author}
  {\bibfnamefont {A.}~\bibnamefont {Kurchanov}}, \bibinfo {author}
  {\bibfnamefont {V.}~\bibnamefont {Luk'yashin}}, \bibinfo {author}
  {\bibfnamefont {V.}~\bibnamefont {Lyashuk}}, \bibinfo {author} {\bibfnamefont
  {E.}~\bibnamefont {Novikov}}, \bibinfo {author} {\bibfnamefont
  {A.}~\bibnamefont {Rostovtsev}}, \bibinfo {author} {\bibfnamefont
  {A.}~\bibnamefont {Sokolov}}, \bibinfo {author} {\bibfnamefont
  {U.}~\bibnamefont {Feizkhanov}},\ and\ \bibinfo {author} {\bibfnamefont
  {N.}~\bibnamefont {Khaldeeva}},\ }\href
  {https://doi.org/10.1023/A:1017520322662} {\bibfield  {journal} {\bibinfo
  {journal} {Instruments and Experimental Techniques}\ }\textbf {\bibinfo
  {volume} {44}},\ \bibinfo {pages} {327 – 334} (\bibinfo {year}
  {2001})}\BibitemShut {NoStop}%
\bibitem [{\citenamefont {{Budnev, N.M.}}\ \emph {et~al.}(2017)\citenamefont
  {{Budnev, N.M.}}, \citenamefont {{Avrorin, A.D.}}, \citenamefont {{Avrorin,
  A.V.}}, \citenamefont {{Aynutdinov, V.M.}}, \citenamefont {{Bannasch, R.}},
  \citenamefont {{Belolaptikov, I.A.}}, \citenamefont {{Bogorodsky, D.Yu.}},
  \citenamefont {{Brudanin, V.B.}}, \citenamefont {{Danilchenko, I.A.}},
  \citenamefont {{Domogatsky, G.V.}}, \citenamefont {{Doroshenko, A.A.}},
  \citenamefont {{Dyachok, A.N.}}, \citenamefont {{Dzhilkibaev, Zh.-A.M.}},
  \citenamefont {{Fialkovsky, S.V.}}, \citenamefont {{Gafarov, A.R.}},
  \citenamefont {{Gaponenko, O.N.}}, \citenamefont {{Golubkov, K.V.}},
  \citenamefont {{Gress, T.I.}}, \citenamefont {{Honz, Z.}}, \citenamefont
  {{Kebkal, K.G.}}, \citenamefont {{Kebkal, O.G.}}, \citenamefont {{Konischev,
  K.V.}}, \citenamefont {{Korobchenko, A.V.}}, \citenamefont {{Koshechkin,
  A.P.}}, \citenamefont {{Koshel, F.K.}}, \citenamefont {{Kozhin, A.V.}},
  \citenamefont {{Kulepov, V.F.}}, \citenamefont {{Kuleshov, D.A.}},
  \citenamefont {{Ljashuk, V.I.}}, \citenamefont {{Milenin, M.B.}},
  \citenamefont {{Mirgazov, R.R.}}, \citenamefont {{Osipova, E.R.}},
  \citenamefont {{Panfilov, A.I}}, \citenamefont {{Pankov, L.V.}},
  \citenamefont {{Perevalov, A.A.}}, \citenamefont {{Pliskovsky, E.N.}},
  \citenamefont {{Rjabov, E.V}}, \citenamefont {{Rozanov, M.I.}}, \citenamefont
  {{Rubtzov, V.Yu.}}, \citenamefont {{Shaybonov, B.A}}, \citenamefont
  {{Sheifler, A.A.}}, \citenamefont {{Shelepov, M.D}}, \citenamefont
  {{Shkurihin, A.V.}}, \citenamefont {{Smagina, A.A.}}, \citenamefont
  {{Suvorova, O.V.}}, \citenamefont {{Tabolenko, V.A}}, \citenamefont
  {{Tarashansky, B.A.}}, \citenamefont {{Yakovlev, S.A.}}, \citenamefont
  {{Zagorodnikov, A.V}},\ and\ \citenamefont {{Zurbanov, V.L.}}}]{Budnev2017}%
  \BibitemOpen
  \bibfield  {author} {\bibinfo {author} {\bibnamefont {{Budnev, N.M.}}},
  \bibinfo {author} {\bibnamefont {{Avrorin, A.D.}}}, \bibinfo {author}
  {\bibnamefont {{Avrorin, A.V.}}}, \bibinfo {author} {\bibnamefont
  {{Aynutdinov, V.M.}}}, \bibinfo {author} {\bibnamefont {{Bannasch, R.}}},
  \bibinfo {author} {\bibnamefont {{Belolaptikov, I.A.}}}, \bibinfo {author}
  {\bibnamefont {{Bogorodsky, D.Yu.}}}, \bibinfo {author} {\bibnamefont
  {{Brudanin, V.B.}}}, \bibinfo {author} {\bibnamefont {{Danilchenko, I.A.}}},
  \bibinfo {author} {\bibnamefont {{Domogatsky, G.V.}}}, \bibinfo {author}
  {\bibnamefont {{Doroshenko, A.A.}}}, \bibinfo {author} {\bibnamefont
  {{Dyachok, A.N.}}}, \bibinfo {author} {\bibnamefont {{Dzhilkibaev,
  Zh.-A.M.}}}, \bibinfo {author} {\bibnamefont {{Fialkovsky, S.V.}}}, \bibinfo
  {author} {\bibnamefont {{Gafarov, A.R.}}}, \bibinfo {author} {\bibnamefont
  {{Gaponenko, O.N.}}}, \bibinfo {author} {\bibnamefont {{Golubkov, K.V.}}},
  \bibinfo {author} {\bibnamefont {{Gress, T.I.}}}, \bibinfo {author}
  {\bibnamefont {{Honz, Z.}}}, \bibinfo {author} {\bibnamefont {{Kebkal,
  K.G.}}}, \bibinfo {author} {\bibnamefont {{Kebkal, O.G.}}}, \bibinfo {author}
  {\bibnamefont {{Konischev, K.V.}}}, \bibinfo {author} {\bibnamefont
  {{Korobchenko, A.V.}}}, \bibinfo {author} {\bibnamefont {{Koshechkin,
  A.P.}}}, \bibinfo {author} {\bibnamefont {{Koshel, F.K.}}}, \bibinfo {author}
  {\bibnamefont {{Kozhin, A.V.}}}, \bibinfo {author} {\bibnamefont {{Kulepov,
  V.F.}}}, \bibinfo {author} {\bibnamefont {{Kuleshov, D.A.}}}, \bibinfo
  {author} {\bibnamefont {{Ljashuk, V.I.}}}, \bibinfo {author} {\bibnamefont
  {{Milenin, M.B.}}}, \bibinfo {author} {\bibnamefont {{Mirgazov, R.R.}}},
  \bibinfo {author} {\bibnamefont {{Osipova, E.R.}}}, \bibinfo {author}
  {\bibnamefont {{Panfilov, A.I}}}, \bibinfo {author} {\bibnamefont {{Pankov,
  L.V.}}}, \bibinfo {author} {\bibnamefont {{Perevalov, A.A.}}}, \bibinfo
  {author} {\bibnamefont {{Pliskovsky, E.N.}}}, \bibinfo {author} {\bibnamefont
  {{Rjabov, E.V}}}, \bibinfo {author} {\bibnamefont {{Rozanov, M.I.}}},
  \bibinfo {author} {\bibnamefont {{Rubtzov, V.Yu.}}}, \bibinfo {author}
  {\bibnamefont {{Shaybonov, B.A}}}, \bibinfo {author} {\bibnamefont
  {{Sheifler, A.A.}}}, \bibinfo {author} {\bibnamefont {{Shelepov, M.D}}},
  \bibinfo {author} {\bibnamefont {{Shkurihin, A.V.}}}, \bibinfo {author}
  {\bibnamefont {{Smagina, A.A.}}}, \bibinfo {author} {\bibnamefont {{Suvorova,
  O.V.}}}, \bibinfo {author} {\bibnamefont {{Tabolenko, V.A}}}, \bibinfo
  {author} {\bibnamefont {{Tarashansky, B.A.}}}, \bibinfo {author}
  {\bibnamefont {{Yakovlev, S.A.}}}, \bibinfo {author} {\bibnamefont
  {{Zagorodnikov, A.V}}},\ and\ \bibinfo {author} {\bibnamefont {{Zurbanov,
  V.L.}}},\ }\href {https://doi.org/10.1051/epjconf/201713506004} {\bibfield
  {journal} {\bibinfo  {journal} {EPJ Web Conf.}\ }\textbf {\bibinfo {volume}
  {135}},\ \bibinfo {pages} {06004} (\bibinfo {year} {2017})}\BibitemShut
  {NoStop}%
\bibitem [{\citenamefont {Ishihara}(2021)}]{Ishihara2019}%
  \BibitemOpen
  \bibfield  {author} {\bibinfo {author} {\bibfnamefont {A.}~\bibnamefont
  {Ishihara}} (\bibinfo {collaboration} {IceCube}),\ }\href
  {https://doi.org/10.22323/1.358.1031} {\bibfield  {journal} {\bibinfo
  {journal} {PoS}\ }\textbf {\bibinfo {volume} {ICRC2019}},\ \bibinfo {pages}
  {1031} (\bibinfo {year} {2021})},\ \Eprint {https://arxiv.org/abs/1908.09441}
  {arXiv:1908.09441 [astro-ph.HE]} \BibitemShut {NoStop}%
\bibitem [{\citenamefont {Nosengo}(2009)}]{Nosengo2009}%
  \BibitemOpen
  \bibfield  {author} {\bibinfo {author} {\bibfnamefont {N.}~\bibnamefont
  {Nosengo}},\ }\href {https://doi.org/10.1038/462560a} {\bibfield  {journal}
  {\bibinfo  {journal} {Nature}\ }\textbf {\bibinfo {volume} {462}},\ \bibinfo
  {pages} {560} (\bibinfo {year} {2009})}\BibitemShut {NoStop}%
\bibitem [{\citenamefont {Baxter}\ \emph {et~al.}(2017)\citenamefont {Baxter},
  \citenamefont {Chen}, \citenamefont {Crisler}, \citenamefont {Cwiok},
  \citenamefont {Dahl}, \citenamefont {Grimsted}, \citenamefont {Gupta},
  \citenamefont {Jin}, \citenamefont {Puig}, \citenamefont {Temples},\ and\
  \citenamefont {Zhang}}]{Baxter2017}%
  \BibitemOpen
  \bibfield  {author} {\bibinfo {author} {\bibfnamefont {D.}~\bibnamefont
  {Baxter}}, \bibinfo {author} {\bibfnamefont {C.~J.}\ \bibnamefont {Chen}},
  \bibinfo {author} {\bibfnamefont {M.}~\bibnamefont {Crisler}}, \bibinfo
  {author} {\bibfnamefont {T.}~\bibnamefont {Cwiok}}, \bibinfo {author}
  {\bibfnamefont {C.~E.}\ \bibnamefont {Dahl}}, \bibinfo {author}
  {\bibfnamefont {A.}~\bibnamefont {Grimsted}}, \bibinfo {author}
  {\bibfnamefont {J.}~\bibnamefont {Gupta}}, \bibinfo {author} {\bibfnamefont
  {M.}~\bibnamefont {Jin}}, \bibinfo {author} {\bibfnamefont {R.}~\bibnamefont
  {Puig}}, \bibinfo {author} {\bibfnamefont {D.}~\bibnamefont {Temples}},\ and\
  \bibinfo {author} {\bibfnamefont {J.}~\bibnamefont {Zhang}},\ }\href
  {https://doi.org/10.1103/PhysRevLett.118.231301} {\bibfield  {journal}
  {\bibinfo  {journal} {Phys. Rev. Lett.}\ }\textbf {\bibinfo {volume} {118}},\
  \bibinfo {pages} {231301} (\bibinfo {year} {2017})}\BibitemShut {NoStop}%
\bibitem [{\citenamefont {Barnabé-Heider}\ \emph {et~al.}(2005)\citenamefont
  {Barnabé-Heider}, \citenamefont {{Di Marco}}, \citenamefont {Doane},
  \citenamefont {Genest}, \citenamefont {Gornea}, \citenamefont {Guénette},
  \citenamefont {Leroy}, \citenamefont {Lessard}, \citenamefont {Martin},
  \citenamefont {Wichoski}, \citenamefont {Zacek}, \citenamefont {Clark},
  \citenamefont {Krauss}, \citenamefont {Noble}, \citenamefont {Behnke},
  \citenamefont {Feighery}, \citenamefont {Levine}, \citenamefont {Muthusi},
  \citenamefont {Kanagalingam},\ and\ \citenamefont {Noulty}}]{Barnabe2005}%
  \BibitemOpen
  \bibfield  {author} {\bibinfo {author} {\bibfnamefont {M.}~\bibnamefont
  {Barnabé-Heider}}, \bibinfo {author} {\bibfnamefont {M.}~\bibnamefont {{Di
  Marco}}}, \bibinfo {author} {\bibfnamefont {P.}~\bibnamefont {Doane}},
  \bibinfo {author} {\bibfnamefont {M.-H.}\ \bibnamefont {Genest}}, \bibinfo
  {author} {\bibfnamefont {R.}~\bibnamefont {Gornea}}, \bibinfo {author}
  {\bibfnamefont {R.}~\bibnamefont {Guénette}}, \bibinfo {author}
  {\bibfnamefont {C.}~\bibnamefont {Leroy}}, \bibinfo {author} {\bibfnamefont
  {L.}~\bibnamefont {Lessard}}, \bibinfo {author} {\bibfnamefont {J.-P.}\
  \bibnamefont {Martin}}, \bibinfo {author} {\bibfnamefont {U.}~\bibnamefont
  {Wichoski}}, \bibinfo {author} {\bibfnamefont {V.}~\bibnamefont {Zacek}},
  \bibinfo {author} {\bibfnamefont {K.}~\bibnamefont {Clark}}, \bibinfo
  {author} {\bibfnamefont {C.}~\bibnamefont {Krauss}}, \bibinfo {author}
  {\bibfnamefont {A.}~\bibnamefont {Noble}}, \bibinfo {author} {\bibfnamefont
  {E.}~\bibnamefont {Behnke}}, \bibinfo {author} {\bibfnamefont
  {W.}~\bibnamefont {Feighery}}, \bibinfo {author} {\bibfnamefont
  {I.}~\bibnamefont {Levine}}, \bibinfo {author} {\bibfnamefont
  {C.}~\bibnamefont {Muthusi}}, \bibinfo {author} {\bibfnamefont
  {S.}~\bibnamefont {Kanagalingam}},\ and\ \bibinfo {author} {\bibfnamefont
  {R.}~\bibnamefont {Noulty}},\ }\href
  {https://doi.org/https://doi.org/10.1016/j.nima.2005.09.015} {\bibfield
  {journal} {\bibinfo  {journal} {Nuclear Instruments and Methods in Physics
  Research Section A: Accelerators, Spectrometers, Detectors and Associated
  Equipment}\ }\textbf {\bibinfo {volume} {555}},\ \bibinfo {pages} {184}
  (\bibinfo {year} {2005})}\BibitemShut {NoStop}%
\bibitem [{\citenamefont {Behnke}\ \emph {et~al.}(2008)\citenamefont {Behnke},
  \citenamefont {Collar}, \citenamefont {Cooper}, \citenamefont {Crum},
  \citenamefont {Crisler}, \citenamefont {Hu}, \citenamefont {Levine},
  \citenamefont {Nakazawa}, \citenamefont {Nguyen}, \citenamefont {Odom},
  \citenamefont {Ramberg}, \citenamefont {Rasmussen}, \citenamefont {Riley},
  \citenamefont {Sonnenschein}, \citenamefont {Szydagis},\ and\ \citenamefont
  {Tschirhart}}]{Behnke2008}%
  \BibitemOpen
  \bibfield  {author} {\bibinfo {author} {\bibfnamefont {E.}~\bibnamefont
  {Behnke}}, \bibinfo {author} {\bibfnamefont {J.~I.}\ \bibnamefont {Collar}},
  \bibinfo {author} {\bibfnamefont {P.~S.}\ \bibnamefont {Cooper}}, \bibinfo
  {author} {\bibfnamefont {K.}~\bibnamefont {Crum}}, \bibinfo {author}
  {\bibfnamefont {M.}~\bibnamefont {Crisler}}, \bibinfo {author} {\bibfnamefont
  {M.}~\bibnamefont {Hu}}, \bibinfo {author} {\bibfnamefont {I.}~\bibnamefont
  {Levine}}, \bibinfo {author} {\bibfnamefont {D.}~\bibnamefont {Nakazawa}},
  \bibinfo {author} {\bibfnamefont {H.}~\bibnamefont {Nguyen}}, \bibinfo
  {author} {\bibfnamefont {B.}~\bibnamefont {Odom}}, \bibinfo {author}
  {\bibfnamefont {E.}~\bibnamefont {Ramberg}}, \bibinfo {author} {\bibfnamefont
  {J.}~\bibnamefont {Rasmussen}}, \bibinfo {author} {\bibfnamefont
  {N.}~\bibnamefont {Riley}}, \bibinfo {author} {\bibfnamefont
  {A.}~\bibnamefont {Sonnenschein}}, \bibinfo {author} {\bibfnamefont
  {M.}~\bibnamefont {Szydagis}},\ and\ \bibinfo {author} {\bibfnamefont
  {R.}~\bibnamefont {Tschirhart}},\ }\href
  {https://doi.org/10.1126/science.1149999} {\bibfield  {journal} {\bibinfo
  {journal} {Science}\ }\textbf {\bibinfo {volume} {319}},\ \bibinfo {pages}
  {933} (\bibinfo {year} {2008})},\ \Eprint
  {https://arxiv.org/abs/https://www.science.org/doi/pdf/10.1126/science.1149999}
  {https://www.science.org/doi/pdf/10.1126/science.1149999} \BibitemShut
  {NoStop}%
\bibitem [{\citenamefont {Amole}\ \emph {et~al.}(2019)\citenamefont {Amole},
  \citenamefont {Ardid}, \citenamefont {Arnquist}, \citenamefont {Asner},
  \citenamefont {Baxter}, \citenamefont {Behnke}, \citenamefont {Bressler},
  \citenamefont {Broerman}, \citenamefont {Cao}, \citenamefont {Chen},
  \citenamefont {Chowdhury}, \citenamefont {Clark}, \citenamefont {Collar},
  \citenamefont {Cooper}, \citenamefont {Coutu}, \citenamefont {Cowles},
  \citenamefont {Crisler}, \citenamefont {Crowder}, \citenamefont
  {Cruz-Venegas}, \citenamefont {Dahl}, \citenamefont {Das}, \citenamefont
  {Fallows}, \citenamefont {Farine}, \citenamefont {Felis}, \citenamefont
  {Filgas}, \citenamefont {Girard}, \citenamefont {Giroux}, \citenamefont
  {Hall}, \citenamefont {Hardy}, \citenamefont {Harris}, \citenamefont
  {Hillier}, \citenamefont {Hoppe}, \citenamefont {Jackson}, \citenamefont
  {Jin}, \citenamefont {Klopfenstein}, \citenamefont {Kozynets}, \citenamefont
  {Krauss}, \citenamefont {Laurin}, \citenamefont {Lawson}, \citenamefont
  {Leblanc}, \citenamefont {Levine}, \citenamefont {Licciardi}, \citenamefont
  {Lippincott}, \citenamefont {Loer}, \citenamefont {Mamedov}, \citenamefont
  {Mitra}, \citenamefont {Moore}, \citenamefont {Nania}, \citenamefont
  {Neilson}, \citenamefont {Noble}, \citenamefont {Oedekerk}, \citenamefont
  {Ortega}, \citenamefont {Piro}, \citenamefont {Plante}, \citenamefont
  {Podviyanuk}, \citenamefont {Priya}, \citenamefont {Robinson}, \citenamefont
  {Sahoo}, \citenamefont {Scallon}, \citenamefont {Seth}, \citenamefont
  {Sonnenschein}, \citenamefont {Starinski}, \citenamefont
  {\ifmmode~\check{S}\else \v{S}\fi{}tekl}, \citenamefont {Sullivan},
  \citenamefont {Tardif}, \citenamefont {V\'azquez-J\'auregui}, \citenamefont
  {Walkowski}, \citenamefont {Weima}, \citenamefont {Wichoski}, \citenamefont
  {Wierman}, \citenamefont {Yan}, \citenamefont {Zacek},\ and\ \citenamefont
  {Zhang}}]{Amole2019}%
  \BibitemOpen
  \bibfield  {author} {\bibinfo {author} {\bibfnamefont {C.}~\bibnamefont
  {Amole}}, \bibinfo {author} {\bibfnamefont {M.}~\bibnamefont {Ardid}},
  \bibinfo {author} {\bibfnamefont {I.~J.}\ \bibnamefont {Arnquist}}, \bibinfo
  {author} {\bibfnamefont {D.~M.}\ \bibnamefont {Asner}}, \bibinfo {author}
  {\bibfnamefont {D.}~\bibnamefont {Baxter}}, \bibinfo {author} {\bibfnamefont
  {E.}~\bibnamefont {Behnke}}, \bibinfo {author} {\bibfnamefont
  {M.}~\bibnamefont {Bressler}}, \bibinfo {author} {\bibfnamefont
  {B.}~\bibnamefont {Broerman}}, \bibinfo {author} {\bibfnamefont
  {G.}~\bibnamefont {Cao}}, \bibinfo {author} {\bibfnamefont {C.~J.}\
  \bibnamefont {Chen}}, \bibinfo {author} {\bibfnamefont {U.}~\bibnamefont
  {Chowdhury}}, \bibinfo {author} {\bibfnamefont {K.}~\bibnamefont {Clark}},
  \bibinfo {author} {\bibfnamefont {J.~I.}\ \bibnamefont {Collar}}, \bibinfo
  {author} {\bibfnamefont {P.~S.}\ \bibnamefont {Cooper}}, \bibinfo {author}
  {\bibfnamefont {C.~B.}\ \bibnamefont {Coutu}}, \bibinfo {author}
  {\bibfnamefont {C.}~\bibnamefont {Cowles}}, \bibinfo {author} {\bibfnamefont
  {M.}~\bibnamefont {Crisler}}, \bibinfo {author} {\bibfnamefont
  {G.}~\bibnamefont {Crowder}}, \bibinfo {author} {\bibfnamefont {N.~A.}\
  \bibnamefont {Cruz-Venegas}}, \bibinfo {author} {\bibfnamefont {C.~E.}\
  \bibnamefont {Dahl}}, \bibinfo {author} {\bibfnamefont {M.}~\bibnamefont
  {Das}}, \bibinfo {author} {\bibfnamefont {S.}~\bibnamefont {Fallows}},
  \bibinfo {author} {\bibfnamefont {J.}~\bibnamefont {Farine}}, \bibinfo
  {author} {\bibfnamefont {I.}~\bibnamefont {Felis}}, \bibinfo {author}
  {\bibfnamefont {R.}~\bibnamefont {Filgas}}, \bibinfo {author} {\bibfnamefont
  {F.}~\bibnamefont {Girard}}, \bibinfo {author} {\bibfnamefont
  {G.}~\bibnamefont {Giroux}}, \bibinfo {author} {\bibfnamefont
  {J.}~\bibnamefont {Hall}}, \bibinfo {author} {\bibfnamefont {C.}~\bibnamefont
  {Hardy}}, \bibinfo {author} {\bibfnamefont {O.}~\bibnamefont {Harris}},
  \bibinfo {author} {\bibfnamefont {T.}~\bibnamefont {Hillier}}, \bibinfo
  {author} {\bibfnamefont {E.~W.}\ \bibnamefont {Hoppe}}, \bibinfo {author}
  {\bibfnamefont {C.~M.}\ \bibnamefont {Jackson}}, \bibinfo {author}
  {\bibfnamefont {M.}~\bibnamefont {Jin}}, \bibinfo {author} {\bibfnamefont
  {L.}~\bibnamefont {Klopfenstein}}, \bibinfo {author} {\bibfnamefont
  {T.}~\bibnamefont {Kozynets}}, \bibinfo {author} {\bibfnamefont {C.~B.}\
  \bibnamefont {Krauss}}, \bibinfo {author} {\bibfnamefont {M.}~\bibnamefont
  {Laurin}}, \bibinfo {author} {\bibfnamefont {I.}~\bibnamefont {Lawson}},
  \bibinfo {author} {\bibfnamefont {A.}~\bibnamefont {Leblanc}}, \bibinfo
  {author} {\bibfnamefont {I.}~\bibnamefont {Levine}}, \bibinfo {author}
  {\bibfnamefont {C.}~\bibnamefont {Licciardi}}, \bibinfo {author}
  {\bibfnamefont {W.~H.}\ \bibnamefont {Lippincott}}, \bibinfo {author}
  {\bibfnamefont {B.}~\bibnamefont {Loer}}, \bibinfo {author} {\bibfnamefont
  {F.}~\bibnamefont {Mamedov}}, \bibinfo {author} {\bibfnamefont
  {P.}~\bibnamefont {Mitra}}, \bibinfo {author} {\bibfnamefont
  {C.}~\bibnamefont {Moore}}, \bibinfo {author} {\bibfnamefont
  {T.}~\bibnamefont {Nania}}, \bibinfo {author} {\bibfnamefont
  {R.}~\bibnamefont {Neilson}}, \bibinfo {author} {\bibfnamefont {A.~J.}\
  \bibnamefont {Noble}}, \bibinfo {author} {\bibfnamefont {P.}~\bibnamefont
  {Oedekerk}}, \bibinfo {author} {\bibfnamefont {A.}~\bibnamefont {Ortega}},
  \bibinfo {author} {\bibfnamefont {M.-C.}\ \bibnamefont {Piro}}, \bibinfo
  {author} {\bibfnamefont {A.}~\bibnamefont {Plante}}, \bibinfo {author}
  {\bibfnamefont {R.}~\bibnamefont {Podviyanuk}}, \bibinfo {author}
  {\bibfnamefont {S.}~\bibnamefont {Priya}}, \bibinfo {author} {\bibfnamefont
  {A.~E.}\ \bibnamefont {Robinson}}, \bibinfo {author} {\bibfnamefont
  {S.}~\bibnamefont {Sahoo}}, \bibinfo {author} {\bibfnamefont
  {O.}~\bibnamefont {Scallon}}, \bibinfo {author} {\bibfnamefont
  {S.}~\bibnamefont {Seth}}, \bibinfo {author} {\bibfnamefont {A.}~\bibnamefont
  {Sonnenschein}}, \bibinfo {author} {\bibfnamefont {N.}~\bibnamefont
  {Starinski}}, \bibinfo {author} {\bibfnamefont {I.}~\bibnamefont
  {\ifmmode~\check{S}\else \v{S}\fi{}tekl}}, \bibinfo {author} {\bibfnamefont
  {T.}~\bibnamefont {Sullivan}}, \bibinfo {author} {\bibfnamefont
  {F.}~\bibnamefont {Tardif}}, \bibinfo {author} {\bibfnamefont
  {E.}~\bibnamefont {V\'azquez-J\'auregui}}, \bibinfo {author} {\bibfnamefont
  {N.}~\bibnamefont {Walkowski}}, \bibinfo {author} {\bibfnamefont
  {E.}~\bibnamefont {Weima}}, \bibinfo {author} {\bibfnamefont
  {U.}~\bibnamefont {Wichoski}}, \bibinfo {author} {\bibfnamefont
  {K.}~\bibnamefont {Wierman}}, \bibinfo {author} {\bibfnamefont
  {Y.}~\bibnamefont {Yan}}, \bibinfo {author} {\bibfnamefont {V.}~\bibnamefont
  {Zacek}},\ and\ \bibinfo {author} {\bibfnamefont {J.}~\bibnamefont {Zhang}}
  (\bibinfo {collaboration} {PICO Collaboration}),\ }\href
  {https://doi.org/10.1103/PhysRevD.100.022001} {\bibfield  {journal} {\bibinfo
   {journal} {Phys. Rev. D}\ }\textbf {\bibinfo {volume} {100}},\ \bibinfo
  {pages} {022001} (\bibinfo {year} {2019})}\BibitemShut {NoStop}%
\bibitem [{\citenamefont {{Amole, C.}}\ \emph {et~al.}(2015)\citenamefont
  {{Amole, C.}}, \citenamefont {{Ardid, M.}}, \citenamefont {{Asner, D. M.}},
  \citenamefont {{Baxter, D.}}, \citenamefont {{Behnke, E.}}, \citenamefont
  {{Bhattacharjee, P.}}, \citenamefont {{Borsodi, H.}}, \citenamefont
  {{Bou-Cabo, M.}}, \citenamefont {{Brice, S. J.}}, \citenamefont
  {{Broemmelsiek, D.}}, \citenamefont {{Clark, K.}}, \citenamefont {{Collar, J.
  I.}}, \citenamefont {{Cooper, P. S.}}, \citenamefont {{Crisler, M.}},
  \citenamefont {{Dahl, C. E.}}, \citenamefont {{Das, M.}}, \citenamefont
  {{Debris, F.}}, \citenamefont {{Dhungana, N.}}, \citenamefont {{Farine, J.}},
  \citenamefont {{Felis, I.}}, \citenamefont {{Filgas, R.}}, \citenamefont
  {{Fines-Neuschild, M.}}, \citenamefont {{Girard, F.}}, \citenamefont
  {{Giroux, G.}}, \citenamefont {{Hai, M.}}, \citenamefont {{Hall, J.}},
  \citenamefont {{Harris, O.}}, \citenamefont {{Jackson, C. M.}}, \citenamefont
  {{Jin, M.}}, \citenamefont {{Krauss, C.}}, \citenamefont {{Lafreni\`ere,
  M.}}, \citenamefont {{Laurin, M.}}, \citenamefont {{Lawson, I.}},
  \citenamefont {{Levine, I.}}, \citenamefont {{Lippincott, W. H.}},
  \citenamefont {{Mann, E.}}, \citenamefont {{Martin, J. P.}}, \citenamefont
  {{Maurya, D.}}, \citenamefont {{Mitra, P.}}, \citenamefont {{Neilson, R.}},
  \citenamefont {{Noble, A. J.}}, \citenamefont {{Plante, A.}}, \citenamefont
  {{Podviyanuk, R.}}, \citenamefont {{Priya, S.}}, \citenamefont {{Robinson, A.
  E.}}, \citenamefont {{Ruschman, M.}}, \citenamefont {{Scallon, O.}},
  \citenamefont {{Seth, S.}}, \citenamefont {{Sonnenschein, A.}}, \citenamefont
  {{Starinski, N.}}, \citenamefont {{Stekl, I.}}, \citenamefont
  {{V\'azquez-J\'auregui, E.}}, \citenamefont {{Wells, J.}}, \citenamefont
  {{Wichoski, U.}}, \citenamefont {{Zacek, V.}},\ and\ \citenamefont {{Zhang,
  J.}}}]{Amole2015}%
  \BibitemOpen
  \bibfield  {author} {\bibinfo {author} {\bibnamefont {{Amole, C.}}}, \bibinfo
  {author} {\bibnamefont {{Ardid, M.}}}, \bibinfo {author} {\bibnamefont
  {{Asner, D. M.}}}, \bibinfo {author} {\bibnamefont {{Baxter, D.}}}, \bibinfo
  {author} {\bibnamefont {{Behnke, E.}}}, \bibinfo {author} {\bibnamefont
  {{Bhattacharjee, P.}}}, \bibinfo {author} {\bibnamefont {{Borsodi, H.}}},
  \bibinfo {author} {\bibnamefont {{Bou-Cabo, M.}}}, \bibinfo {author}
  {\bibnamefont {{Brice, S. J.}}}, \bibinfo {author} {\bibnamefont
  {{Broemmelsiek, D.}}}, \bibinfo {author} {\bibnamefont {{Clark, K.}}},
  \bibinfo {author} {\bibnamefont {{Collar, J. I.}}}, \bibinfo {author}
  {\bibnamefont {{Cooper, P. S.}}}, \bibinfo {author} {\bibnamefont {{Crisler,
  M.}}}, \bibinfo {author} {\bibnamefont {{Dahl, C. E.}}}, \bibinfo {author}
  {\bibnamefont {{Das, M.}}}, \bibinfo {author} {\bibnamefont {{Debris, F.}}},
  \bibinfo {author} {\bibnamefont {{Dhungana, N.}}}, \bibinfo {author}
  {\bibnamefont {{Farine, J.}}}, \bibinfo {author} {\bibnamefont {{Felis,
  I.}}}, \bibinfo {author} {\bibnamefont {{Filgas, R.}}}, \bibinfo {author}
  {\bibnamefont {{Fines-Neuschild, M.}}}, \bibinfo {author} {\bibnamefont
  {{Girard, F.}}}, \bibinfo {author} {\bibnamefont {{Giroux, G.}}}, \bibinfo
  {author} {\bibnamefont {{Hai, M.}}}, \bibinfo {author} {\bibnamefont {{Hall,
  J.}}}, \bibinfo {author} {\bibnamefont {{Harris, O.}}}, \bibinfo {author}
  {\bibnamefont {{Jackson, C. M.}}}, \bibinfo {author} {\bibnamefont {{Jin,
  M.}}}, \bibinfo {author} {\bibnamefont {{Krauss, C.}}}, \bibinfo {author}
  {\bibnamefont {{Lafreni\`ere, M.}}}, \bibinfo {author} {\bibnamefont
  {{Laurin, M.}}}, \bibinfo {author} {\bibnamefont {{Lawson, I.}}}, \bibinfo
  {author} {\bibnamefont {{Levine, I.}}}, \bibinfo {author} {\bibnamefont
  {{Lippincott, W. H.}}}, \bibinfo {author} {\bibnamefont {{Mann, E.}}},
  \bibinfo {author} {\bibnamefont {{Martin, J. P.}}}, \bibinfo {author}
  {\bibnamefont {{Maurya, D.}}}, \bibinfo {author} {\bibnamefont {{Mitra,
  P.}}}, \bibinfo {author} {\bibnamefont {{Neilson, R.}}}, \bibinfo {author}
  {\bibnamefont {{Noble, A. J.}}}, \bibinfo {author} {\bibnamefont {{Plante,
  A.}}}, \bibinfo {author} {\bibnamefont {{Podviyanuk, R.}}}, \bibinfo {author}
  {\bibnamefont {{Priya, S.}}}, \bibinfo {author} {\bibnamefont {{Robinson, A.
  E.}}}, \bibinfo {author} {\bibnamefont {{Ruschman, M.}}}, \bibinfo {author}
  {\bibnamefont {{Scallon, O.}}}, \bibinfo {author} {\bibnamefont {{Seth,
  S.}}}, \bibinfo {author} {\bibnamefont {{Sonnenschein, A.}}}, \bibinfo
  {author} {\bibnamefont {{Starinski, N.}}}, \bibinfo {author} {\bibnamefont
  {{Stekl, I.}}}, \bibinfo {author} {\bibnamefont {{V\'azquez-J\'auregui,
  E.}}}, \bibinfo {author} {\bibnamefont {{Wells, J.}}}, \bibinfo {author}
  {\bibnamefont {{Wichoski, U.}}}, \bibinfo {author} {\bibnamefont {{Zacek,
  V.}}},\ and\ \bibinfo {author} {\bibnamefont {{Zhang, J.}}},\ }\href
  {https://doi.org/10.1051/epjconf/20149504020} {\bibfield  {journal} {\bibinfo
   {journal} {EPJ Web of Conferences}\ }\textbf {\bibinfo {volume} {95}},\
  \bibinfo {pages} {04020} (\bibinfo {year} {2015})}\BibitemShut {NoStop}%
\bibitem [{\citenamefont {Hunter}(1981)}]{Hunter1981}%
  \BibitemOpen
  \bibfield  {author} {\bibinfo {author} {\bibfnamefont {S.~D.}\ \bibnamefont
  {Hunter}},\ }\bibfield  {journal} {\bibinfo  {journal} {LSU Historical
  Dissertation and Theses}\ }\href
  {https://doi.org/10.31390/gradschool\_disstheses.3639}
  {10.31390/gradschool\_disstheses.3639} (\bibinfo {year} {1981})\BibitemShut
  {NoStop}%
\bibitem [{\citenamefont {Learned}(1979)}]{Learned1979}%
  \BibitemOpen
  \bibfield  {author} {\bibinfo {author} {\bibfnamefont {J.~G.}\ \bibnamefont
  {Learned}},\ }\href {https://doi.org/10.1103/PhysRevD.19.3293} {\bibfield
  {journal} {\bibinfo  {journal} {Physical Review D}\ }\textbf {\bibinfo
  {volume} {19}},\ \bibinfo {pages} {3293} (\bibinfo {year}
  {1979})}\BibitemShut {NoStop}%
\bibitem [{\citenamefont {Bortolan}\ and\ \citenamefont
  {Carvalho}(2015)}]{Bortolan2015}%
  \BibitemOpen
  \bibfield  {author} {\bibinfo {author} {\bibfnamefont {M.~C.}\ \bibnamefont
  {Bortolan}}\ and\ \bibinfo {author} {\bibfnamefont {A.~N.}\ \bibnamefont
  {Carvalho}},\ }\href {https://doi.org/10.12775/TMNA.2015.059} {\bibfield
  {journal} {\bibinfo  {journal} {Topological Methods in Nonlinear Analysis}\
  }\textbf {\bibinfo {volume} {46}},\ \bibinfo {pages} {563 } (\bibinfo {year}
  {2015})}\BibitemShut {NoStop}%
\bibitem [{\citenamefont {Viggen}(2011)}]{Viggen2011}%
  \BibitemOpen
  \bibfield  {author} {\bibinfo {author} {\bibfnamefont {E.~M.}\ \bibnamefont
  {Viggen}},\ }\href {https://doi.org/10.1098/rsta.2011.0040} {\bibfield
  {journal} {\bibinfo  {journal} {Philosophical Transactions of the Royal
  Society A: Mathematical, Physical and Engineering Sciences}\ }\textbf
  {\bibinfo {volume} {369}},\ \bibinfo {pages} {2246} (\bibinfo {year}
  {2011})}\BibitemShut {NoStop}%
\bibitem [{\citenamefont {Settnes}\ and\ \citenamefont
  {Bruus}(2012)}]{Settnes2012}%
  \BibitemOpen
  \bibfield  {author} {\bibinfo {author} {\bibfnamefont {M.}~\bibnamefont
  {Settnes}}\ and\ \bibinfo {author} {\bibfnamefont {H.}~\bibnamefont
  {Bruus}},\ }\href {https://doi.org/10.1103/PhysRevE.85.016327} {\bibfield
  {journal} {\bibinfo  {journal} {Phys. Rev. E}\ }\textbf {\bibinfo {volume}
  {85}},\ \bibinfo {pages} {016327} (\bibinfo {year} {2012})}\BibitemShut
  {NoStop}%
\bibitem [{\citenamefont {Groom}\ and\ \citenamefont
  {Klein}(2021)}]{Groom2021}%
  \BibitemOpen
  \bibfield  {author} {\bibinfo {author} {\bibfnamefont {D.}~\bibnamefont
  {Groom}}\ and\ \bibinfo {author} {\bibfnamefont {S.}~\bibnamefont {Klein}},\
  }\bibinfo {title} {Passage of particles through matter},\ in\ \href@noop {}
  {\emph {\bibinfo {booktitle} {Review of Particle Physics 2021}}},\ Vol.\
  \bibinfo {volume} {2021}\ (\bibinfo  {publisher} {Particle Data Group},\
  \bibinfo {year} {2021})\ pp.\ \bibinfo {pages} {389 -- 412}\BibitemShut
  {NoStop}%
\bibitem [{\citenamefont {Kutcher}\ and\ \citenamefont
  {Green}(1976)}]{Kutcher1976}%
  \BibitemOpen
  \bibfield  {author} {\bibinfo {author} {\bibfnamefont {G.~J.}\ \bibnamefont
  {Kutcher}}\ and\ \bibinfo {author} {\bibfnamefont {A.~E.~S.}\ \bibnamefont
  {Green}},\ }\href {http://www.jstor.org/stable/3574338} {\bibfield  {journal}
  {\bibinfo  {journal} {Radiation Research}\ }\textbf {\bibinfo {volume}
  {67}},\ \bibinfo {pages} {408} (\bibinfo {year} {1976})}\BibitemShut
  {NoStop}%
\bibitem [{\citenamefont {Vandenbroucke}\ \emph {et~al.}(2005)\citenamefont
  {Vandenbroucke}, \citenamefont {Gratta},\ and\ \citenamefont
  {Lehtinen}}]{Vandenbroucke2005}%
  \BibitemOpen
  \bibfield  {author} {\bibinfo {author} {\bibfnamefont {J.}~\bibnamefont
  {Vandenbroucke}}, \bibinfo {author} {\bibfnamefont {G.}~\bibnamefont
  {Gratta}},\ and\ \bibinfo {author} {\bibfnamefont {N.}~\bibnamefont
  {Lehtinen}},\ }\href {https://doi.org/10.1086/425336} {\bibfield  {journal}
  {\bibinfo  {journal} {The Astrophysical Journal}\ }\textbf {\bibinfo {volume}
  {621}},\ \bibinfo {pages} {301–312} (\bibinfo {year} {2005})}\BibitemShut
  {NoStop}%
\bibitem [{\citenamefont {Morse}\ and\ \citenamefont
  {Ingard}(1986)}]{Morse1986}%
  \BibitemOpen
  \bibfield  {author} {\bibinfo {author} {\bibfnamefont {P.}~\bibnamefont
  {Morse}}\ and\ \bibinfo {author} {\bibfnamefont {K.}~\bibnamefont {Ingard}},\
  }\href {https://books.google.com/books?id=KIL4MV9IE5kC} {\emph {\bibinfo
  {title} {Theoretical Acoustics}}},\ International series in pure and applied
  physics\ (\bibinfo  {publisher} {Princeton University Press},\ \bibinfo
  {year} {1986})\BibitemShut {NoStop}%
\bibitem [{\citenamefont {Lyamshev}(2004)}]{Lyamshev2004}%
  \BibitemOpen
  \bibfield  {author} {\bibinfo {author} {\bibfnamefont {L.}~\bibnamefont
  {Lyamshev}},\ }\href {https://books.google.com/books?id=3Lj450hL2YcC} {\emph
  {\bibinfo {title} {Radiation Acoustics}}}\ (\bibinfo  {publisher} {CRC
  Press},\ \bibinfo {year} {2004})\BibitemShut {NoStop}%
\bibitem [{\citenamefont {Amsler}(2010)}]{Amsler2010}%
  \BibitemOpen
  \bibfield  {author} {\bibinfo {author} {\bibfnamefont {C.}~\bibnamefont
  {Amsler}},\ }in\ \href {https://doi.org/10.1017/cbo9780511707179.015} {\emph
  {\bibinfo {booktitle} {REVIEW OF PARTICLE PHYSICS}}},\ Vol.~\bibinfo {volume}
  {1}\ (\bibinfo  {publisher} {Particle Data Group},\ \bibinfo {year} {2010})\
  pp.\ \bibinfo {pages} {411--450}\BibitemShut {NoStop}%
\bibitem [{\citenamefont {Lahmann}(2016)}]{Lahmann2016}%
  \BibitemOpen
  \bibfield  {author} {\bibinfo {author} {\bibfnamefont {R.}~\bibnamefont
  {Lahmann}},\ }\href {https://doi.org/10.1016/j.nuclphysbps.2015.09.059}
  {\bibfield  {journal} {\bibinfo  {journal} {Nuclear and Particle Physics
  Proceedings}\ }\textbf {\bibinfo {volume} {273-275}},\ \bibinfo {pages} {406}
  (\bibinfo {year} {2016})}\BibitemShut {NoStop}%
\bibitem [{\citenamefont {Huber}\ \emph {et~al.}(2009)\citenamefont {Huber},
  \citenamefont {Perkins}, \citenamefont {Laesecke}, \citenamefont {Friend},
  \citenamefont {Sengers}, \citenamefont {Assael}, \citenamefont {Metaxa},
  \citenamefont {Vogel}, \citenamefont {Mareš},\ and\ \citenamefont
  {Miyagawa}}]{Huber2009}%
  \BibitemOpen
  \bibfield  {author} {\bibinfo {author} {\bibfnamefont {M.~L.}\ \bibnamefont
  {Huber}}, \bibinfo {author} {\bibfnamefont {R.~A.}\ \bibnamefont {Perkins}},
  \bibinfo {author} {\bibfnamefont {A.}~\bibnamefont {Laesecke}}, \bibinfo
  {author} {\bibfnamefont {D.~G.}\ \bibnamefont {Friend}}, \bibinfo {author}
  {\bibfnamefont {J.~V.}\ \bibnamefont {Sengers}}, \bibinfo {author}
  {\bibfnamefont {M.~J.}\ \bibnamefont {Assael}}, \bibinfo {author}
  {\bibfnamefont {I.~N.}\ \bibnamefont {Metaxa}}, \bibinfo {author}
  {\bibfnamefont {E.}~\bibnamefont {Vogel}}, \bibinfo {author} {\bibfnamefont
  {R.}~\bibnamefont {Mareš}},\ and\ \bibinfo {author} {\bibfnamefont
  {K.}~\bibnamefont {Miyagawa}},\ }\href {https://doi.org/10.1063/1.3088050}
  {\bibfield  {journal} {\bibinfo  {journal} {Journal of Physical and Chemical
  Reference Data}\ }\textbf {\bibinfo {volume} {38}},\ \bibinfo {pages} {101}
  (\bibinfo {year} {2009})},\ \Eprint
  {https://arxiv.org/abs/https://doi.org/10.1063/1.3088050}
  {https://doi.org/10.1063/1.3088050} \BibitemShut {NoStop}%
\bibitem [{\citenamefont {Wagner}\ and\ \citenamefont
  {Pruß}(2002)}]{Wagner2002}%
  \BibitemOpen
  \bibfield  {author} {\bibinfo {author} {\bibfnamefont {W.}~\bibnamefont
  {Wagner}}\ and\ \bibinfo {author} {\bibfnamefont {A.}~\bibnamefont {Pruß}},\
  }\href {https://doi.org/10.1063/1.1461829} {\bibfield  {journal} {\bibinfo
  {journal} {Journal of Physical and Chemical Reference Data}\ }\textbf
  {\bibinfo {volume} {31}},\ \bibinfo {pages} {387} (\bibinfo {year} {2002})},\
  \Eprint {https://arxiv.org/abs/https://doi.org/10.1063/1.1461829}
  {https://doi.org/10.1063/1.1461829} \BibitemShut {NoStop}%
\bibitem [{\citenamefont {Lemmon}\ \emph {et~al.}(2022)\citenamefont {Lemmon},
  \citenamefont {Bell}, \citenamefont {Huber},\ and\ \citenamefont
  {McLinden}}]{Lemmon2022}%
  \BibitemOpen
  \bibfield  {author} {\bibinfo {author} {\bibfnamefont {E.~W.}\ \bibnamefont
  {Lemmon}}, \bibinfo {author} {\bibfnamefont {I.~H.}\ \bibnamefont {Bell}},
  \bibinfo {author} {\bibfnamefont {M.~L.}\ \bibnamefont {Huber}},\ and\
  \bibinfo {author} {\bibfnamefont {M.~O.}\ \bibnamefont {McLinden}},\ }in\
  \href@noop {} {\emph {\bibinfo {booktitle} {NIST Chemistry WebBook}}},\
  \bibinfo {series} {NIST Standard Reference Database}, Vol.~\bibinfo {volume}
  {69},\ \bibinfo {editor} {edited by\ \bibinfo {editor} {\bibfnamefont
  {P.~J.}\ \bibnamefont {Linstrom}}\ and\ \bibinfo {editor} {\bibfnamefont
  {W.~G.}\ \bibnamefont {Mallard}}}\ (\bibinfo  {publisher} {National Institute
  of Standards and Technology},\ \bibinfo {year} {2022})\ Chap.\ \bibinfo
  {chapter} {Thermophysical Properties of Fluid Systems}\BibitemShut {NoStop}%
\bibitem [{\citenamefont {Sharma}\ \emph {et~al.}(2013)\citenamefont {Sharma},
  \citenamefont {Bhagour}, \citenamefont {Sharma},\ and\ \citenamefont
  {Solanki}}]{Sharma2013}%
  \BibitemOpen
  \bibfield  {author} {\bibinfo {author} {\bibfnamefont {V.~K.}\ \bibnamefont
  {Sharma}}, \bibinfo {author} {\bibfnamefont {S.}~\bibnamefont {Bhagour}},
  \bibinfo {author} {\bibfnamefont {D.}~\bibnamefont {Sharma}},\ and\ \bibinfo
  {author} {\bibfnamefont {S.}~\bibnamefont {Solanki}},\ }\href
  {https://doi.org/10.1016/J.TCA.2013.04.011} {\bibfield  {journal} {\bibinfo
  {journal} {Thermochimica Acta}\ }\textbf {\bibinfo {volume} {563}},\ \bibinfo
  {pages} {72} (\bibinfo {year} {2013})}\BibitemShut {NoStop}%
\bibitem [{\citenamefont {Groom}(2021)}]{Groom2021b}%
  \BibitemOpen
  \bibfield  {author} {\bibinfo {author} {\bibfnamefont {D.}~\bibnamefont
  {Groom}},\ }in\ \href {https://pdg.lbl.gov/2022/AtomicNuclearProperties/}
  {\emph {\bibinfo {booktitle} {Review of Particle Physics}}},\ \bibinfo
  {editor} {edited by\ \bibinfo {editor} {\bibfnamefont {R.}~\bibnamefont
  {Workman}} \emph {et~al.}}\ (\bibinfo  {publisher} {Particle Data Group},\
  \bibinfo {year} {2021})\ Chap.\ \bibinfo {chapter} {Atomic and Nuclear
  Properties},\ \bibinfo {note} {to be published (2022)},\ \Eprint
  {https://arxiv.org/abs/https://pdg.lbl.gov/2022/AtomicNuclearProperties/}
  {https://pdg.lbl.gov/2022/AtomicNuclearProperties/} \BibitemShut {NoStop}%
\bibitem [{\citenamefont {Inui}\ \emph {et~al.}(2005)\citenamefont {Inui},
  \citenamefont {Ishikawa}, \citenamefont {Matsuda}, \citenamefont {Tamura},
  \citenamefont {Tsutsui},\ and\ \citenamefont {Baron}}]{Inui2005}%
  \BibitemOpen
  \bibfield  {author} {\bibinfo {author} {\bibfnamefont {M.}~\bibnamefont
  {Inui}}, \bibinfo {author} {\bibfnamefont {D.}~\bibnamefont {Ishikawa}},
  \bibinfo {author} {\bibfnamefont {K.}~\bibnamefont {Matsuda}}, \bibinfo
  {author} {\bibfnamefont {K.}~\bibnamefont {Tamura}}, \bibinfo {author}
  {\bibfnamefont {S.}~\bibnamefont {Tsutsui}},\ and\ \bibinfo {author}
  {\bibfnamefont {A.}~\bibnamefont {Baron}},\ }\href
  {https://doi.org/https://doi.org/10.1016/j.jpcs.2005.09.021} {\bibfield
  {journal} {\bibinfo  {journal} {Journal of Physics and Chemistry of Solids}\
  }\textbf {\bibinfo {volume} {66}},\ \bibinfo {pages} {2223} (\bibinfo {year}
  {2005})},\ \bibinfo {note} {5th International Conference on Inelastic X-ray
  Scattering (IXS 2004)}\BibitemShut {NoStop}%
\bibitem [{\citenamefont {Singh}\ \emph {et~al.}(2007)\citenamefont {Singh},
  \citenamefont {Arafin},\ and\ \citenamefont {George}}]{Singh2007}%
  \BibitemOpen
  \bibfield  {author} {\bibinfo {author} {\bibfnamefont {R.}~\bibnamefont
  {Singh}}, \bibinfo {author} {\bibfnamefont {S.}~\bibnamefont {Arafin}},\ and\
  \bibinfo {author} {\bibfnamefont {A.}~\bibnamefont {George}},\ }\href
  {https://doi.org/https://doi.org/10.1016/j.physb.2006.04.029} {\bibfield
  {journal} {\bibinfo  {journal} {Physica B: Condensed Matter}\ }\textbf
  {\bibinfo {volume} {387}},\ \bibinfo {pages} {344} (\bibinfo {year}
  {2007})}\BibitemShut {NoStop}%
\bibitem [{\citenamefont {Habashi}(2013)}]{Habashi2013}%
  \BibitemOpen
  \bibfield  {author} {\bibinfo {author} {\bibfnamefont {F.}~\bibnamefont
  {Habashi}},\ }\bibinfo {title} {Mercury, physical and chemical properties},\
  in\ \href {https://doi.org/10.1007/978-1-4614-1533-6_307} {\emph {\bibinfo
  {booktitle} {Encyclopedia of Metalloproteins}}},\ \bibinfo {editor} {edited
  by\ \bibinfo {editor} {\bibfnamefont {R.~H.}\ \bibnamefont {Kretsinger}},
  \bibinfo {editor} {\bibfnamefont {V.~N.}\ \bibnamefont {Uversky}},\ and\
  \bibinfo {editor} {\bibfnamefont {E.~A.}\ \bibnamefont {Permyakov}}}\
  (\bibinfo  {publisher} {Springer New York},\ \bibinfo {address} {New York,
  NY},\ \bibinfo {year} {2013})\ pp.\ \bibinfo {pages} {1375--1377}\BibitemShut
  {NoStop}%
\bibitem [{\citenamefont {Davis}\ and\ \citenamefont
  {Gordon}(1967)}]{Davis1967}%
  \BibitemOpen
  \bibfield  {author} {\bibinfo {author} {\bibfnamefont {L.~A.}\ \bibnamefont
  {Davis}}\ and\ \bibinfo {author} {\bibfnamefont {R.~B.}\ \bibnamefont
  {Gordon}},\ }\href {https://doi.org/10.1063/1.1841095} {\bibfield  {journal}
  {\bibinfo  {journal} {The Journal of Chemical Physics}\ }\textbf {\bibinfo
  {volume} {46}},\ \bibinfo {pages} {2650} (\bibinfo {year} {1967})},\ \Eprint
  {https://arxiv.org/abs/https://doi.org/10.1063/1.1841095}
  {https://doi.org/10.1063/1.1841095} \BibitemShut {NoStop}%
\bibitem [{\citenamefont {Rao}\ and\ \citenamefont {Dutta}(1983)}]{Rao1983}%
  \BibitemOpen
  \bibfield  {author} {\bibinfo {author} {\bibfnamefont {R.~V.~G.}\
  \bibnamefont {Rao}}\ and\ \bibinfo {author} {\bibfnamefont {S.~K.}\
  \bibnamefont {Dutta}},\ }\href {https://doi.org/doi:10.1515/zpch-1983-26493}
  {\bibfield  {journal} {\bibinfo  {journal} {Zeitschrift für Physikalische
  Chemie}\ }\textbf {\bibinfo {volume} {264O}},\ \bibinfo {pages} {771}
  (\bibinfo {year} {1983})}\BibitemShut {NoStop}%
\bibitem [{\citenamefont {Lemmon}\ and\ \citenamefont
  {Jacobsen}(2004)}]{Lemmon2004}%
  \BibitemOpen
  \bibfield  {author} {\bibinfo {author} {\bibfnamefont {E.~W.}\ \bibnamefont
  {Lemmon}}\ and\ \bibinfo {author} {\bibfnamefont {R.~T.}\ \bibnamefont
  {Jacobsen}},\ }\href {https://doi.org/10.1023/B:IJOT.0000022327.04529.f3}
  {\bibfield  {journal} {\bibinfo  {journal} {International Journal of
  Thermophysics}\ }\textbf {\bibinfo {volume} {25}},\ \bibinfo {pages} {21}
  (\bibinfo {year} {2004})}\BibitemShut {NoStop}%
\bibitem [{\citenamefont {Tegeler}\ \emph {et~al.}(1999)\citenamefont
  {Tegeler}, \citenamefont {Span},\ and\ \citenamefont {Wagner}}]{Tegeler1999}%
  \BibitemOpen
  \bibfield  {author} {\bibinfo {author} {\bibfnamefont {C.}~\bibnamefont
  {Tegeler}}, \bibinfo {author} {\bibfnamefont {R.}~\bibnamefont {Span}},\ and\
  \bibinfo {author} {\bibfnamefont {W.}~\bibnamefont {Wagner}},\ }\href
  {https://doi.org/10.1063/1.556037} {\bibfield  {journal} {\bibinfo  {journal}
  {Journal of Physical and Chemical Reference Data}\ }\textbf {\bibinfo
  {volume} {28}},\ \bibinfo {pages} {779} (\bibinfo {year} {1999})},\ \Eprint
  {https://arxiv.org/abs/https://doi.org/10.1063/1.556037}
  {https://doi.org/10.1063/1.556037} \BibitemShut {NoStop}%
\bibitem [{\citenamefont {Span}\ \emph {et~al.}(2000)\citenamefont {Span},
  \citenamefont {Lemmon}, \citenamefont {Jacobsen}, \citenamefont {Wagner},\
  and\ \citenamefont {Yokozeki}}]{Span2000}%
  \BibitemOpen
  \bibfield  {author} {\bibinfo {author} {\bibfnamefont {R.}~\bibnamefont
  {Span}}, \bibinfo {author} {\bibfnamefont {E.~W.}\ \bibnamefont {Lemmon}},
  \bibinfo {author} {\bibfnamefont {R.~T.}\ \bibnamefont {Jacobsen}}, \bibinfo
  {author} {\bibfnamefont {W.}~\bibnamefont {Wagner}},\ and\ \bibinfo {author}
  {\bibfnamefont {A.}~\bibnamefont {Yokozeki}},\ }\href
  {https://doi.org/10.1063/1.1349047} {\bibfield  {journal} {\bibinfo
  {journal} {Journal of Physical and Chemical Reference Data}\ }\textbf
  {\bibinfo {volume} {29}},\ \bibinfo {pages} {1361} (\bibinfo {year}
  {2000})},\ \Eprint {https://arxiv.org/abs/https://doi.org/10.1063/1.1349047}
  {https://doi.org/10.1063/1.1349047} \BibitemShut {NoStop}%
\bibitem [{\citenamefont {Lemmon}\ and\ \citenamefont
  {Span}(2006)}]{Lemmon2006}%
  \BibitemOpen
  \bibfield  {author} {\bibinfo {author} {\bibfnamefont {E.~W.}\ \bibnamefont
  {Lemmon}}\ and\ \bibinfo {author} {\bibfnamefont {R.}~\bibnamefont {Span}},\
  }\href {https://doi.org/10.1021/je050186n} {\bibfield  {journal} {\bibinfo
  {journal} {Journal of Chemical \& Engineering Data}\ }\textbf {\bibinfo
  {volume} {51}},\ \bibinfo {pages} {785} (\bibinfo {year} {2006})},\ \Eprint
  {https://arxiv.org/abs/https://doi.org/10.1021/je050186n}
  {https://doi.org/10.1021/je050186n} \BibitemShut {NoStop}%
\bibitem [{\citenamefont {Huber}(2018)}]{Huber2018}%
  \BibitemOpen
  \bibfield  {author} {\bibinfo {author} {\bibfnamefont {M.}~\bibnamefont
  {Huber}},\ }\href {https://doi.org/https://doi.org/10.6028/NIST.IR.8209}
  {\bibinfo {title} {Models for viscosity, thermal conductivity, and surface
  tension of selected pure fluids as implemented in refprop v10.0}} (\bibinfo
  {year} {2018})\BibitemShut {NoStop}%
\bibitem [{\citenamefont {Sternheimer}(1961)}]{Sternheimer1961}%
  \BibitemOpen
  \bibfield  {author} {\bibinfo {author} {\bibfnamefont {R.~M.}\ \bibnamefont
  {Sternheimer}},\ }\href {https://doi.org/10.1016/S0076-695X(08)60442-1}
  {\bibfield  {journal} {\bibinfo  {journal} {Methods in Experimental Physics}\
  }\textbf {\bibinfo {volume} {5}},\ \bibinfo {pages} {1} (\bibinfo {year}
  {1961})}\BibitemShut {NoStop}%
\bibitem [{\citenamefont {Oikonomou}\ \emph {et~al.}(2021)\citenamefont
  {Oikonomou}, \citenamefont {Arneodo},\ and\ \citenamefont
  {Manenti}}]{Oikonomou2021}%
  \BibitemOpen
  \bibfield  {author} {\bibinfo {author} {\bibfnamefont {P.}~\bibnamefont
  {Oikonomou}}, \bibinfo {author} {\bibfnamefont {F.}~\bibnamefont {Arneodo}},\
  and\ \bibinfo {author} {\bibfnamefont {L.}~\bibnamefont {Manenti}},\
  }\bibfield  {journal} {\bibinfo  {journal} {GitHub Repository}\ }\href
  {https://doi.org/doi:https://github.com/PanosEconomou/LXe-Phonon}
  {doi:https://github.com/PanosEconomou/LXe-Phonon} (\bibinfo {year}
  {2021})\BibitemShut {NoStop}%
\end{thebibliography}%

\end{document}